\documentclass[12pt,aps,pre,article,floatfix,showpacs,showkeys]{revtex4-1}

\usepackage{times}
\usepackage{amsmath}    
\usepackage{amsfonts,bm}   
\usepackage{amssymb}
\usepackage{graphicx}
\usepackage{epsfig}
\usepackage{subfig}
\usepackage{overpic}

\newcommand{\E}{\mathbb{E}}

\begin{document}

\title{A Stochastic Stick - Slip Model Linking Crustal Shear Strength and  Earthquake Interevent Times}

\author{Dionissios T. Hristopulos}
 \email{dionisi@mred.tuc.gr}   
 \homepage[web: ]{www.mred.tuc.gr/home/hristopoulos/dionisi.htm}
 \affiliation{Technical University of Crete, Geostatistics Research Unit,
Chania 73100, Greece}
\author{Vasiliki Mouslopoulou}
 \email{vasiliki@mred.tuc.gr}
 \affiliation{Technical University of Crete, Geostatistics Research Unit,
Chania 73100, Greece}

\date{\today}

\begin{abstract}
\begin{description}
\item[Background] The current understanding of the \textit{earthquake interevent times distribution} (ITD) in terms of seismological laws or fundamental physical principles is incomplete. The Weibull distribution is used to model the earthquake ITD.
\item[Purpose] To better understand the earthquake ITD  in terms of  fracture mechanics.
\item[Method] We link the earthquake ITD on single faults with the Earth's crustal
 shear strength distribution, for which we use the Weibull model, by means of a phenomenological, stochastic stick - slip model. We generalize the model for fault systems.
\item[Results] We obtain Weibull  ITD  for  power-law stress accumulation, i.e., $\sigma(t) = \alpha \, t^{\beta}$, where $\beta >0$ for single faults or fault systems with homogeneous strength statistics.   We also show that logarithmic stress accumulation leads to the log-Weibull ITD. For the Weibull ITD, we prove that (i) $m= \beta \, m_s$, where $m$ and $m_s$ are, respectively, the interevent times and crustal shear strength Weibull moduli  and (ii) the  time scale
 $\tau_s = (S_s/\alpha)^{1/\beta}$ where $S_s$ is the  scale of crustal shear strength. We investigate deviations of the ITD tails from the Weibull due to sampling bias, magnitude selection, and non-homogeneous strength parameters.   Assuming the Gutenberg - Richter law and independence of $m$ on the magnitude threshold,  $M_{L,c},$ we deduce that  $\tau_s \propto e^{- \rho_{M} M_{L,c}},$ where
$\rho_M \in [1.15, 3.45]$ for seismically active regions. We demonstrate that a microearthquake
sequence conforms reasonably well to the Weibull model.
\item[Conclusions] The stochastic stick - slip model justifies the Weibull  ITD for single faults and for homogeneous fault systems, while it suggests mixtures of Weibull distributions for heterogeneous fault systems. Non-universal deviations from Weibull statistics are  possible, even for single faults, due to magnitude thresholds and non-uniform parameter values.
\end{description}
\end{abstract}



\maketitle

\newpage


\section{INTRODUCTION}

Tectonic earthquakes are the result of the stick - slip motion of plates within the
Earth's crust. This motion can be viewed in the framework of
driven dissipative systems.  Earthquakes are complex processes that involve nonlinearities, stochasticity, and multiple spatiotemporal scales. Earthquakes originate on individual faults, which can be viewed as
the fundamental structural units for earthquake processes. Faults are geological structures
that span a variety of scales, from a few centimeters up to thousands of kilometers
for plate boundary faults, and accommodate earthquakes that span up to eight orders of magnitude
in size~\cite{Scholz02,Sornette99}. Neighboring faults interact with each other leading to space-time organization of earthquakes within \textit{fault systems}, e.g.~\cite{MH11}. The statistical laws of seismicity, e.g., the Gutenberg - Richter scaling law, are valid for
\textit{systems of faults} but not necessarily for single faults~\cite{Serino11}.

One of the most interesting problems in statistical seismology and statistical  physics is
the probability distribution of the times between
successive earthquake events that exceed a specified magnitude threshold, i.e.,
the so-called \textit{interevent times
distribution}. The terms
\textit{return intervals, waiting times and recurrence
intervals} are also used instead of interevent times.  A subtle but important distinction can
be made between \textit{recurrence intervals}, which refer to
consecutive events that take place on the same fault, and
\textit{interocurrence intervals}, which focus on all faults in a
specified region~\cite{Abaimov08}. The statistical properties of recurrence
intervals are more difficult to estimate, because less information is available
for individual faults. The distinction is, nevertheless, conceptually important, since
recurrence intervals characterize the \emph{one-body}, i.e., the single-fault, problem, while
interoccurrence intervals are associated with the activity of the \emph{many-body system}~\cite{Sornette99}.
In the following we use the term interevent times in both cases, but we distinguish between models that apply to the one-body versus the many-body problems.

Knowledge of the interevent times distribution is important for the
assessment of earthquake hazard in seismically active regions. From a practical perspective,
inferring interevent times statistics for very large earthquakes based on the
statistics of smaller-magnitude and higher-frequency events is highly desirable.
In seismology, earthquakes are categorized as foreshocks, main shocks and aftershocks.
Aftershocks  are assumed to be triggered by dynamic stress changes induced by the main event, such as  stress redistribution,
 fluid flow triggering, etc., that are not directly linked to the tectonic motion of the plates.
 The classification of earthquakes as main events, foreshocks and aftershocks (declustering),
 is based on heuristic arguments. The  result of declustering analysis strongly depends on the
  method used~\cite{Luen11}. Given this ambiguity and the fact that all seismic events
 in a fault system can be viewed as the result of the same dynamic process, we do not distinguish between different types of events; this  approach is commonly used in statistical physics studies~\cite{Bak02,Corral03,Corral04,Corral06a,Saichev07}.

\subsection{Notation}

We briefly comment on some notation and abbreviations used in the following: the probability density function  of a random variable $X(t)$ is denoted by $f_{\rm X}(x)$ and the cumulative probability function by $F_{\rm X}(x)=P(X\le x)$. The strength of the Earth's  crust is denoted by the random variable  $S$, which is assumed to follow Weibull statistics with shape parameter $m_s$ and scale parameter $S_{s}.$   The  local magnitude (Richter scale) of an earthquake event will be denoted  by $M_L$, and the magnitude threshold  by the parameter $M_{L,c}$~\footnote{We use the local magnitude instead of the moment magnitude since the earthquake sequence we investigate involves small to moderate size earthquakes.}. The earthquake times are denoted by the random variable $t_{i}$, where the index $i$ counts the natural time (i.e. the event order), and the interevent times are denoted by $\tau_{i}=t_{i+1}-t_{i}$. Finally, $\hat{A}$  will denote the estimate of $A$ from data, where $A$ stands for a function or a distribution parameter.
We also use the following abbreviations: \emph{PDF} for the probability density function, \emph{CDF} for the cumulative distribution function, \emph{ITD} for the interevent times distribution and \emph{CSS} for the Cretan seismic sequence.

\subsection{Earthquake interevent times}
The distributions of the magnitudes and return intervals of extreme events
are research topics that attract significant interest~\cite{sornette04a}.
Extreme events are usually distributed in
both space and time, but  herein we assume
that the spatial dependence can be ignored to simplify the analysis. Let us further assume
that an \textit{extreme event} corresponds to excursions of a random
function $X(t)$, where $t$ is the time index, to values above a
specified threshold $z_q$. The classical extreme value theory
focuses on the probability distributions of such extreme
events~\cite{Gumbel35}. The
Fisher-Tippet-Gnedenko theorem~\cite{Fisher28,Gnedenko43}  shows
 that the distribution of $M_n := \max
(X_1, \ldots, X_n)$, where the $X_i, i=1,\ldots,n$ are independent
and identically distributed (i.i.d.) variables, is given by 
Generalized Extreme Value (GEV) distributions. The GEV include the Gumbel, reverse Weibull
and Fr\'{e}chet distributions. The type of
GEV obtained depends on the tail behavior of the probability
distribution of the $X_i$. Similar distributions, but with reversed supports, are
obtained for minima.
Extension of the classical CLTs (which
involve deterministic scalings of the random variables), to
randomized CLTs that enable  stochastic scaling transformations have been
recently proposed in~\cite{Klafter10}.

If the earthquakes in a fault system  occurred
randomly, the event times  should be uniformly distributed.
Uniform distribution over an unbounded time domain
 implies \textit{Poisson statistics} as
lucidly explained in~\cite{Klafter10}. The Poisson model  leads
to an exponential distribution of interevent times. It has been proposed that the
earthquake times follow the Poisson distribution if foreshocks and aftershocks are removed~\cite{Gardner74}. Different declustering approaches have been proposed  to isolate
main events from aftershocks and foreshocks. Nevertheless, these approaches are not based
on fundamental principles, and the results of their application on seismic data vary widely.
In addition, a recent rigorous statistical analysis casts doubt on the validity of the Poisson model
even for declustered data~\cite{Luen11}.  The
\textit{periodic model} supports
that the intervals between characteristic earthquakes are approximately constant~\cite{Schwartz84}. This is
 in contrast with the observations collected from large faults globally~\cite{Abaimov08,Marco96,Weldon04}.

Spatial and temporal inter-dependence of seismic
events can explain deviations from the Poisson and the
periodic models.  Several  research papers
propose that earthquakes are self-organized systems, perhaps  near a critical point~\cite{Bak02,Corral03,Saichev07} or systems near a spinodal critical point~\cite{Klein97,Serino11}.
Both cases imply the emergence of power laws in the system.
Bak {\it et al.}~\cite{Bak02} introduced a global scaling law that
relates earthquake interevent intervals with magnitude and distance
between earthquakes. These authors analyzed seismic catalogue data from an
extended area in California
 that includes several faults over a period of 16 years (ca. $3.35\times 10^5$ events).
 They observed power-law dependence over 8 orders of
magnitude, indicating correlations over a wide range of interevent
intervals, distances and magnitudes. Corral and
coworkers~\cite{Corral03,Corral04,Corral06a,Corral06b} introduced a
local modification of the scaling form so that the interevent time
\textit{probability density function} (PDF) follows the universal expression $f_{\rm \tau}(\tau) \simeq \lambda
\tilde{f}(\lambda \, \tau)$, where $\tilde{f}( \tau)$ is a scaling
function and the typical interevent time $\bar{\tau}$ is specific to
the region of interest.

Saichev and Sornette~\cite{Sornette06,Saichev07} generalized the scaling function
 incorporating parameters with locally varying values.
Their analysis was based on
the mean-field approximation of the interevent times PDF in the
epidemic-type aftershock sequence (ETAS) model. ETAS  incorporates  the Gutenberg-Richter dependence of frequency on
magnitude, the Omori - Utsu law for the rate of aftershocks, and  a
similarity assumption that does not  distinguish
between foreshocks, main events and aftershocks (any event can
be considered as a trigger for subsequent events).

In a different research vein, several studies of earthquake catalogues and simulations show
that the Weibull distribution provides a better match of empirical interevent times distributions than the Poisson model~\cite{Hagiwara74,Rikitake76,Rikitake91,Sieh89,Yakovlev06,Abaimov08,Hasumi09a,Hasumi09b}.
The arguments supporting the \textit{Weibull distribution} are based on empirical
studies and extreme value theory.  In particular, since the interevent times are
determined by minima of the shear strength in the Earth's crust, the standard Weibull model is a good candidate
for their distribution. In contrast,
the Gumbel distribution for minima has negative support and the Fr\'{e}chet distribution has
an unbounded support.

\subsection{Aims and outline of this paper}
In this paper we propose a \textit{stochastic stick - slip model} that links the shear strength (in the following referred to as strength for brevity)
distribution of faults in the Earth's crust, the stress accumulation - relaxation process in the crust
due to  tectonic motion, and the earthquakes interevent times distribution. The model is formulated at the single-fault scale and is then extended to a system of faults by constructing a composite strength distribution.

A prototype physical model of stick - slip motion \textit{along single faults} is the Burridge-Knopoff (BK) model~\cite{BK67,Carlson94,Xia08}.
 This model consists
  of a system of coupled differential equations representing the motion of $n$ point masses linked with elastic springs and subject to a velocity-dependent friction force, which is responsible for the slip instability. In contrast to the BK model, the stochastic stick - slip model proposed herein is phenomenological, since the time between events is determined by a heuristic stress accumulation function.
   We assume that the main stochastic component is due to the variations of the fault shear strength  (or the strength across different faults in a system). Stochastic aspects of the stress accumulation function are not explicitly investigated in the following, but they obviously deserve further research.

The remainder of the paper is organized as follows: In section~\ref{sec:weibull} we  review
the Weibull distribution and its  applications to earthquake interevent times. In Section~\ref{sec:weaklink} we
 propose a stochastic stick - slip model for single faults and show that it admits interevent times
 distributions, including the Weibull, that depend on the time evolution of stress accumulation.
 In particular, we show that the Weibull is an admissible  interevent times model, if (i) the crust strength distribution is Weibull with stationary and homogeneous parameters,
 and (ii) the stress accumulation increases with time as a power-law.
 We also show that deviations from the Weibull can result due to spatial non-homogeneity of the crustal strength parameters and by imposing finite magnitude thresholds on the seismic sequence. In section~\ref{sec:ITD-system} we generalize the model to interevent times for a system of faults.
In Section~\ref{sec:Crete} we investigate the ITD
for a microearthquake  sequence from the island of Crete (Greece) in relation to the proposed stick - slip model.  Finally,  Section~\ref{sec:conclusions} involves a discussion,  conclusions, and topics for further research.

\section{The Weibull distribution and earthquake interevent times  }
\label{sec:weibull}

\subsection{Properties of the Weibull distribution}
The Weibull CDF, $F_{\rm \tau}(\tau)$, determining the probability
that the time between two consecutive events is less than or equal to $\tau \ge 0$ is given by
\begin{equation}
\label{eq:Weib-cdf}
F_{\rm \tau}(\tau) = 1 - \exp\left[ - \left( \frac{\tau}{\tau_{s}} \right)^{m}\right],
\end{equation}
where $\tau_s$ is the \textit{scale parameter}  and $m>0$ is the
\textit{Weibull modulus or shape parameter.}
The PDF is defined by $ f_{\rm \tau}(\tau)=dF_{\rm \tau}(\tau)/d\tau$ and is given by the equation
\begin{equation}
\label{eq:Weib-pdf}
f_{\rm \tau}(\tau) = \frac{m}{\tau_s} \, \left( \frac{\tau}{\tau_{s}} \right)^{m-1}
\, e^{ - \left( \frac{\tau}{\tau_{s}} \right)^{m}}.
\end{equation}
The \emph{survival function} is the complementary cumulative probability function, i.e.,
\begin{equation}
R(\tau) =1 - F_{\rm \tau}(\tau).
\end{equation}
In the Weibull case, $ R(\tau)$  is the \textit{stretched exponential} function $ R(\tau)=e^{- \left( \frac{\tau}{\tau_{s}} \right)^{m}}$.
 The function $R(\tau)$ represents the probability that no seismic event has occurred within time $\tau$  since
 the last event.
Shape parameter values $m<1$ lead to a diverging  density at $\tau=0$, and an almost exponential decay of $ f_{\rm \tau}(\tau)$ as $\tau$ increases.  For $m>1$ the
PDF develops a single peak that becomes sharper with
increasing $m$.

The \textit{hazard rate or failure rate} is the
rate of change for the probability of a seismic event, if time $t$
has elapsed since the last event. It is expressed by
\begin{equation}
\label{eq:hazard}
H(\tau) = \frac{f_{\rm \tau}(\tau)}{1-F_{\rm \tau}(\tau)} =
\frac{m}{\tau_s}\, \left( \frac{\tau}{\tau_s} \right)^{m-1}.
\end{equation}

Data are graphically tested for Weibull dependence using the Weibull plot. The latter employs the fact that the double logarithm of the inverse survival function,
\begin{equation}
\label{eq:Phi}
\Phi(\tau) := \log\log R^{-1}(\tau),
\end{equation}
 satisfies the straight line
equation, $ \Phi(\tau) = m\, \log(\tau) - m \, \log(\tau_s) $ with slope equal to the Weibull modulus.
If the data are drawn from the Weibull distribution, $ \hat{\Phi}(\tau)$ obtained from the empirical CDF, $\hat{F}_{\tau}(\tau)$, is approximately a straight line.

\subsection{The Weibull interevent times distribution}
 The Weibull distribution was investigated in~\cite{Hagiwara74,Rikitake76,Rikitake91}
 for large earthquakes at six
subduction zones over the globe. In~\cite{Sieh89} the ITD of a sequence of 12
paleoearthquakes on the San Andreas fault was investigated. Yakovlev
{\it et al.}~\cite{Yakovlev06}  simulated million-year-long
catalogues of earthquakes on major strike-slip faults in California.
They found that the  Weibull distribution fits the interevent
times better than the lognormal and inverse Gaussian distributions.

Abaimov {\it et al.}~\cite{Abaimov07,Abaimov08} concluded that the Weibull is
a good model for real and simulated
large-magnitude earthquakes on the San Andreas fault, and for a
microearthquake sequence at a nearby site. These authors also
emphasize the  behavior of the hazard rate function of the Weibull
distribution~\cite{Davis89,Sornette97,Corral05a}. At least for large-magnitude
earthquakes,  $H(\tau)$ is expected to increase with the
interevent time. The exponential distribution has a constant $H(\tau)$ indicating lack of memory between events, while for  the lognormal
distribution $H(\tau)$ decreases with the interevent time. The
Brownian-passage time distribution $H(\tau)$ tends to a constant with
increasing interevent time. Of the various distributions considered as
models of earthquake interevent times, an increasing $H(\tau)$ with time
is exhibited only by the Weibull and  gamma
distributions (if $m>1)$. In
addition, Abaimov {\it et al.}~\cite{Abaimov08} find evidence for
the Weibull distribution in numerical solutions of the slider-block
 model introduced by Burridge and Knopoff~\cite{BK67,Langer89}.

Robinson {\it et al.}~\cite{Robinson09} simulated approximately $ 500\ 000$ earthquakes of
magnitudes between 3.8Mw and 6.6Mw (moment magnitude scale), over a period of two million years
for faults in the Taupo Rift in New Zealand. These authors employed a
synthetic seismicity model that is based on the Coulomb failure
criterion and uses empirical data pertaining to the number of faults,
fault lengths, and long-term slip rates. They found that a
three-parameter Weibull distribution  fits the interevent times
of large earthquakes on most faults. The three-parameter Weibull survival function is given by
$R(\tau)= e^{-(\tau - \tau_0)^m/\tau_{s}^m},$ where $\tau_0$ is the \emph{location parameter}.
A finite  $\tau_0$
implies that  the PDF vanish for $\tau < \tau_0$.
According to the analysis in~\cite{Robinson09},
earthquakes within a normal fault system are correlated on a rift-wide scale over time periods
of $\approx 3 \mbox{kyrs}$.

Santhanam and Gantz~\cite{Santhanam08} investigated the
\textit{return intervals of a random function} $X(t)$, i.e.,  the times
between consecutive excursions of
$X(t)$ above a given threshold. They found that if  $X(t)$ has long-range memory (i.e.,
power-law decay of correlations), the return intervals follow the Weibull
distribution. Their mathematical formulation provides support for the Weibull model.
Nevertheless, their approach is not based on
the standard laws of seismology (i.e., Gutenberg-Richter and Omori's law),
and the  values of $m$ admitted are
 $0<m \le 1.$  In contrast, analysis of interevent times from seismic catalogues and simulations
 demonstrates that values $m>1$ also occur. In particular, for large earthquakes it is
 believed that the hazard rate function increases with the time since the last event,
 implying  $m>1.$

\begin{widetext}
\begingroup
\squeezetable
\begin{table*}[htbp]
  \centering
  \caption{Weibull modulus $(m)$ values for earthquake ITD from published research based on empirical and synthetic data from various tectonic settings and system sizes. The magnitudes in column 4 are based on the moment magnitude scale (Mw)~\cite{Kanamori79}. }
    \begin{ruledtabular}
    \begin{tabular}{ccccccccc}
     $\#$ of events & $m$ & Time span & Magnitude (Mw) & Type of data  & System ``Size''& Location  & Reference \\
     7     & 3,21  & 147 yrs & $>6$    & Real & Single Fault & SAF\footnote{SAF: San Andreas Fault, California.}/ Pallett Creek & \cite{Abaimov08} \\
     13    & 2,01  & $\approx$2000 yrs & $>7$    & Real & Single Fault & SAF/Wrightwood & \cite{Biasi02} \\
     13    & 2,91  & $\approx$5 years & Mean=1.36\footnote{Analysis of data from~\cite{Nadeau95}.} & Real & Single Fault & SAF/San Juan Batista &  \cite{Abaimov08} \\
     4606  & 1,97  & 1 Ma\footnote{1Ma=1 Million years} & $>7.5$  & Simulation & Fault system & North SAF / S.S. faults & \cite{Yakovlev06} \\
     5093  & 1,87  & 1 Ma & $>7.5$  & Simulation & Fault system & South SAF  & \cite{Yakovlev06} \\
     2612  & 1,71  & 1 Ma & $>7$    & Simulation & Fault system & Hayward / California & \cite{Yakovlev06} \\
     8174  & 1,42  & 1 Ma & $>6.8$  & Simulation & Fault system & Calaveras / California & \cite{Yakovlev06} \\
     1913  & 1,32  & 1 Ma & $>6.7 $ & Simulation & Fault system & San Gabriel / California & \cite{Yakovlev06} \\
     1075  & 1,7   & 1 Ma & $>7.2$  & Simulation & Fault system & Calaveras / California & \cite{Yakovlev06} \\
     7     & 2,9   & 147 yrs & $>6$    & Real & Fault system & SAF / Parkfield &  \cite{USGS03} \\
     7     & 1,5   & $\approx$ 2000 yrs & $>7$    & Real & Fault system & SAF/ Pallett Creek & \cite{Sieh89} \\
     $5\,10^5$ & 0.82-10\footnote{In~\cite{Robinson09} the data are fitted to the three-parameter Weibull distribution.} & 2 Ma & 3.3-6.8 & Simulation & Fault system & Taupo Rift / New Zealand & \cite{Robinson09} \\
     12024 & 0,91  & 82 months & $>4 $   & Real & Area\footnote{This refers to large areas that may include several fault systems and subduction margins.}
       & Okinawa/Japan & \cite{Hasumi09b} \\
     13678 & 0,79  & 82 months & $>4$    & Real & Area  & Chuetsu/Japan & \cite{Hasumi09b} \\
     12024 & 1,09  & 82 months & $>3.5$  & Real & Area  & Okinawa/Japan & \cite{Hasumi09b} \\
     13678 & 0,85  & 82 months & $>3.5$  & Real & Area  & Chuetsu/Japan & \cite{Hasumi09b} \\
     12024 & 1,43  & 82 months & $>3$    & Real & Area  & Okinawa/Japan & \cite{Hasumi09b} \\
     13678 & 1,08  & 82 months & $>3$    & Real & Area  & Chuetsu/Japan & \cite{Hasumi09b} \\
     12024 & 1,75  & 82 months & $>2$    & Real & Area  & Okinawa/Japan & \cite{Hasumi09b} \\
     13678 & 1,77  & 82 months & $>2$    & Real & Area  & Chuetsu/Japan & \cite{Hasumi09b} \\
           & 1,7   &   70 years    & 6.9-8.4 & Real  & Area  & Japan & \cite{Hagiwara74} \\
     8     & 2,3   & 1000 yrs & 7.9 -8.4 & Real & Subduction margin & Nankai Trench / Japan & \cite{Rikitake76} \\
     16    & 2,9   & 300 yrs & 7.8 -8.4 & Real & Subduction margin & Hokkaido-Kurille Trench / Japan & \cite{Rikitake76} \\
     11    & 9,5   & 100 yrs & 7.8 -8.7 & Real & Subduction margin & Aluetian Trench / Alaska & \cite{Rikitake76} \\
   \end{tabular}
    \end{ruledtabular}
  \label{tab:Weibull}
\end{table*}
\endgroup
\end{widetext}



Table~\ref{tab:Weibull} reviews  published estimates of the Weibull parameters for the interevent times of various earthquake sequences, along with information pertaining to the magnitude, the size of the fault system, the location, and the duration of the
seismic sequence. Note that values of $m>1$ prevail. In addition, estimates of $m$
obtained from records containing a small number of events tend to be higher
 than those from large sample sizes that lie in the range
$0.79 - 1.97$. We believe that this
tendency is partly due to the impact of the magnitude threshold on the ITD
(also see section~\ref{ssec:crust-strength} below). Also note that the data cover very different
system sizes, from single faults to areas containing many fault systems.

\section{A stochastic stick - slip model for single faults}
\label{sec:weaklink}

An overview of the physical processes associated with earthquakes is given by
Kanamori and Brodsky~\cite{Kanamori01,Kanamori04}. We develop our formalism based on  this conceptual framework. Specifically, we establish connections between the shear strength distribution of the Earth's crust, the dynamic process driving earthquake generation, and the distribution of interevent times.
Tectonic earthquakes occur on fault planes due to a dynamic stick-slip process that locally accumulates stress
 caused by the plate motion. This stress accumulation eventually leads to crust failure and stress relaxation when the local crustal strength is exceeded.
The phase of \textit{stress accumulation} (loading phase) is followed by a phase of rapid \textit{stress relaxation}, cf. Fig.3(b) in~\cite{Kanamori04}, which corresponds to the slip events. The stress accumulation - relaxation process is cyclically repeated. This model has been applied with constant stress accumulation rate and constant maximum, $\sigma_{\max},$ or residual, $\sigma_{\rm res},$ stress to explain the recurrence of large earthquakes. If both the maximum and residual stress are constant, the model predicts periodic behavior; if the initial stress is constant, the model predicts the time of the next event; if the final stress is constant, the model predicts that the longer the interevent time, the larger the magnitude of the following event~\cite{Shimazaki80}.

\subsection{On the distribution of crustal strength}
\label{ssec:crust-strength}
Earthquakes are typically localized on faults in the Earth's crust; hence, the strength of these
structures and the applied tectonic loading determine the interevent times on single faults. Since the crust is composed of brittle material (rock), its strength is expected to follow the Weibull probability distribution~\cite{Bazant09}.
The Weibull distribution was derived in the framework of weakest-link theory, founded in the studies of Gumbel and Weibull on extreme-value statistics~\cite{Chakrabarti97}. This theory addresses the strength of brittle and quasibrittle materials  in terms of the strengths of ``representative volume elements'' (RVEs) or
 links~\cite{Curtin98,Hristopulos04,Alava06,Pang08,Alava09}. The material
fails if the RVE with the lowest  strength  breaks, hence the term ``weakest-link''.
The concept of links is straightforward in simple systems,
e.g., one-dimensional chains and fibers. In the case of more complicated structures,
the links represent
critical units that control the failure of the entire system.

Experimental studies on the strength of geological materials, such as rock (granite) under various types of loading (compressive, bending) provide evidence for the validity of the Weibull distribution~\cite{Gupta98,Amaral08}. At the same time,  experimental measurements show that the
crustal strength has a systematic dependence on the depth $h$~\cite{Sibson74,Zoback93,Zoback01}.
Hence, for a single fault the following Weibull CDF is a useful approximation of the strength distribution at depth $h$:
\begin{equation}
\label{eq:CDF-fault-strength}
F_{\rm S}[S;S_{s}(h)] = 1 - e^{- \left( \frac{S}{S_{s}(h)} \right)^{m_s} },
\end{equation}
where $S_s(h)$ is the \textit{strength scale} and $m_s$ is the \emph{strength Weibull modulus}.
Typical values of $m_s$ for laboratory measurements of rock  samples range between 3 and 30~\cite{Gupta98,Amaral08}.

The seismic events on a fault are distributed across different depths in the crust. The systematic dependence of crustal strength measurements on depth~\cite{Aldersons03,Zoback93,Zoback01} implies that
$S_s(h)$ increases continuously with $h$. This agrees qualitatively with the decreasing number of seismic events registered with increasing depth (c.f. Section~\ref{sec:Crete} below).

 The depth-averaged \textit{effective fault strength distribution}  is then given by
\begin{equation}
\label{eq:CDF-fault-eff-strength-int}
F_{\rm S}^{\ast}(S) = 1 - \frac{1}{h_{2} - h_{1}}\int_{h_{1}}^{h_{2}} dh \, e^{- \left( \frac{S}{S_{s}(h)} \right)^{m_s} },
\end{equation}
where $h_1$ is the minimum and $h_2$ the maximum focal depth of the earthquake events.
In Appendix~\ref{App:renorm-weibull} we evaluate the effective distribution for a linear dependence
$S_{s}(h)=S_0+ q\, h$, where $S_{s}(h_{1})=S_0$ and $S_{s}(h_{2})=S_0 + q\, h_{2}$.
It is shown that
  \begin{equation}
  \label{eq:CDF-fault-dav-strength}
 F_{\rm S}^{\ast}(s) =  1-  \frac{1}{2 m_{s} \, \lambda_s }\int_{u_{2}}^{u_{1}} du \, u^{-(1 +1/m_{s})} \, e^{- b\,u}, \quad b=s^{m_s},
\end{equation}
where  $ s = S/\bar{S}$ is the \textit{normalized strength}, $\delta h =2( h_2 - h_1)$,  $\lambda_s= q \, \delta h/\bar{S}$ is the \textit{strength variability coefficient}, and $u_{1,2} = \left(  \frac{1}{1 \mp \lambda_s}\right)^{m_{s}}.$ In Appendix~\ref{App:renorm-weibull} we also derive an explicit expression for the integral valid for ${m_s} \ge 1$ and a convergent infinite series that is valid for  ${m_s}<1$.

In Fig.~\ref{fig:Weibull_correction} we show the numerically integrated CDF $F_{\rm S}^{\ast}(s)$, as well its difference from the Weibull CDF, $F_{0}(s) = 1 - e^{-s^{m_{s}}}$, i.e.,
$\Delta F_{\rm S}^{\ast}(s) = F_{\rm S}^{\ast}(s) - F_{0}(s)$. Larger $m_s$ values lead to
more pronounced  $\Delta F_{\rm S}^{\ast}(s).$ Weibull plots of the effective strength distribution, shown in Fig.~\ref{fig:Phi_effective}, reveal that for $m_s<1$ the linear dependence of $\Phi^{\ast}_{\rm s}(s)= \log\log\frac{1}{1-F_{\rm S}^{\ast}(s)}$ is maintained after the averaging, while $\Phi^{\ast}_{\rm s}(s)$ deviates
 from the  linear (Weibull) dependence for $m_s>1.$
\begin{figure*}
\includegraphics[width=0.9\textwidth]{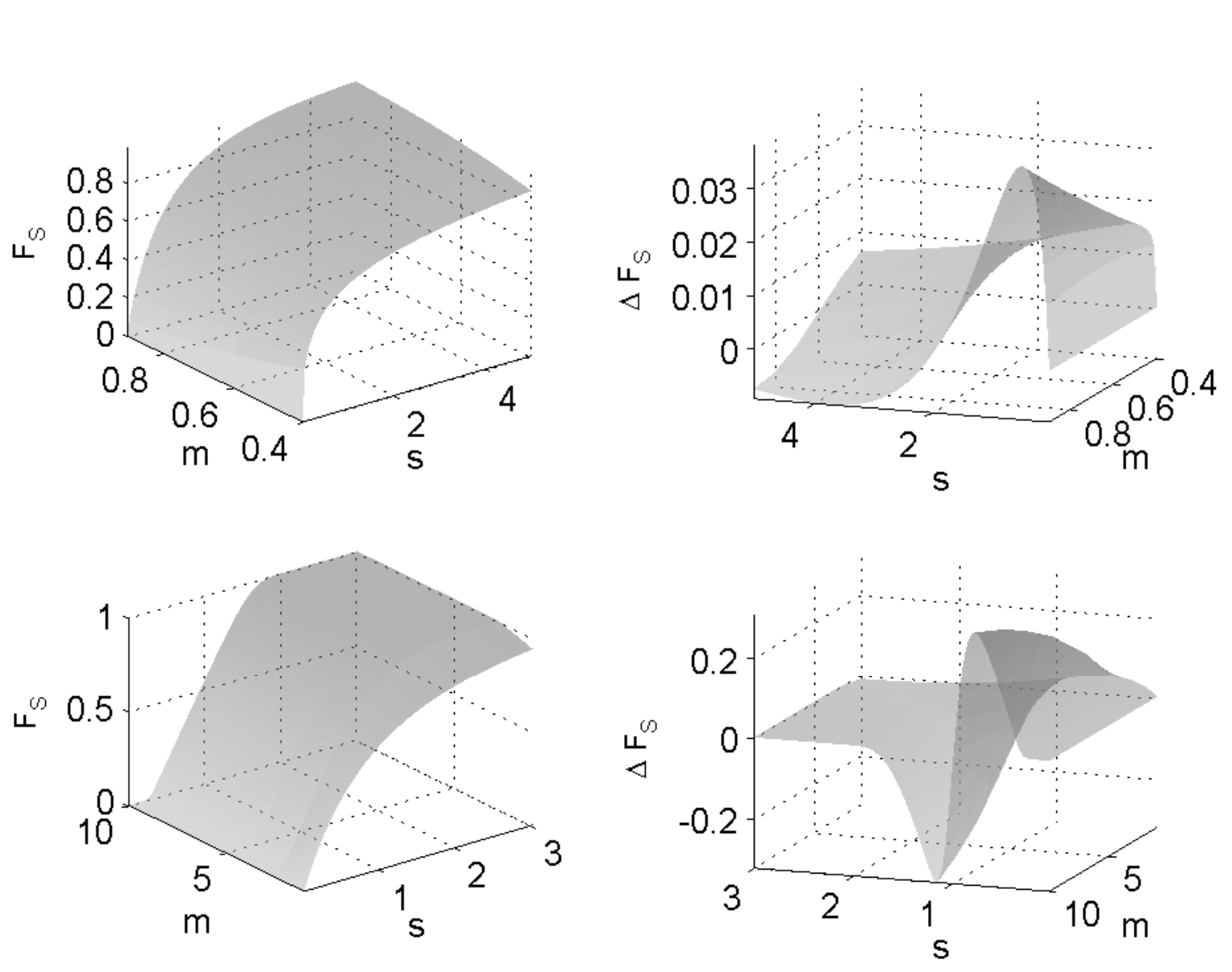}
\caption{\label{fig:Weibull_correction} $F_{\rm S}^{\ast}(s)$ (top left)  and
$\Delta F_{\rm S}^{\ast}(s)$ (top right) for $\lambda_s= 0.7$ and  $0<m_s<1$.  $F_{\rm S}^{\ast}(s)$ (bottom left)  and
$\Delta F_{\rm S}^{\ast}(s)$ (bottom right) for $\lambda_s= 0.7$ and  $10 \ge m_s \ge 1$.}
\label{fig:Weibull_correction}
\end{figure*}

\begin{figure*}
\centering
\subfloat[$\Phi(s)$ for $0 < m_{s} <1$.]{\includegraphics[width=0.47\textwidth]{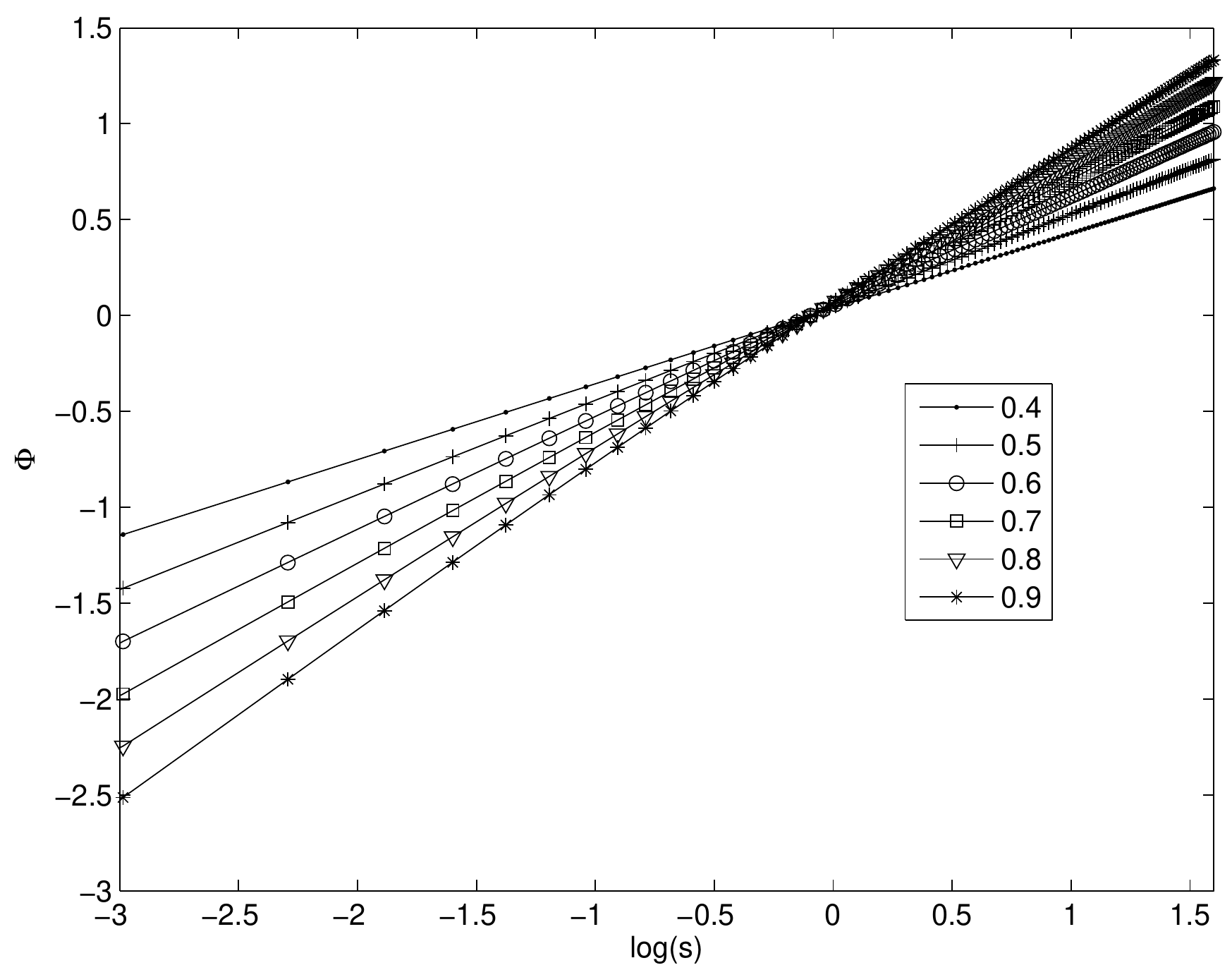}}
\subfloat[$\Phi(s)$ for $10 \ge m_{s} \ge 1$.]{\includegraphics[width=0.47\textwidth]{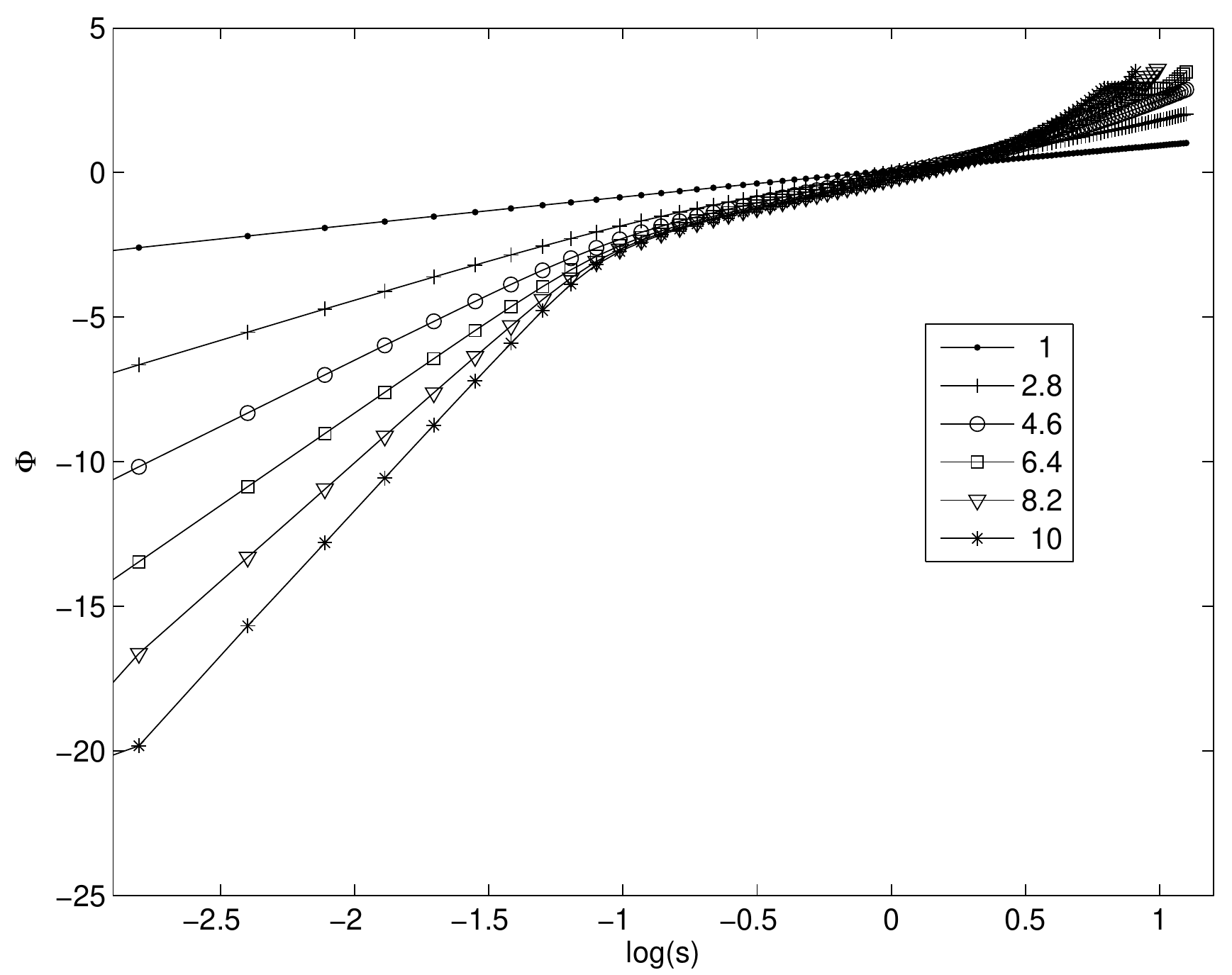}}
\caption{Weibull plots of the effective fault strength distribution based on Eq.~\eqref{eq:CDF-fault-dav-strength}.}
\label{fig:Phi_effective}
\end{figure*}

\subsection{Relation between crustal strength and ITD}
\label{ssec:crust-strength}

The \textit{stochastic stick - slip} model we propose incorporates \textit{non-uniformity} of the crustal strength and links the stress accumulation and relaxation processes with the strength distribution
and  the earthquake interevent times.

Let us assume that $t_{i} \in [0,\tilde{t}_{i}] $ measures the time during the i-th loading phase, and that the stress increase is determined by
$\sigma(t_{i}) = \phi(t_{i}),$ where the \textit{loading function} $\phi(t)$ is an increasing function of time. In general, the parameters of
$\phi(t)$ vary randomly between different phases. For simplicity, we first assume that $\phi(t)$  is a deterministic function, the relaxation time is negligible, and the stress relaxes to a zero residual value after a seismic event. Let the random variable
$\sigma(\tilde{t}_{i})$ correspond to the
stress at failure for the seismic event $E_{i}$. Hence, $\sigma_{\max,i}:=\sigma(\tilde{t}_{i})$ is equal to the strength, $S_{i}$, of the crust at the particular location and time, and we can assume that it follows the Weibull distribution. The time of failure  is given by $\tilde{t}_{i}= \phi^{-1}(S_{i})$, where $\phi^{-1}(\cdot)$ represents the inverse of $\phi(t)$. Once the event $E_{i}$ takes place, the stress is relaxed  and the next event $E_{i+1}$  occurs when  $S_{i+1}=\sigma(\tilde{t}_{i+1})$.  The interevent time between events $E_{i}$ and $E_{i+1}$  is given by $\tilde{t}_{i+1} = \phi^{-1}(S_{i+1})$. In general, if the crustal shear strength $S$ is viewed as a random variable,
it is related to the interevent times random variable, $\tau$, by means of $S=\phi(\tau).$

If $f_{\rm S}(S)$ is the PDF of the crust strength,  and $\phi(\tau)$ is a differentiable, monotone function, the corresponding PDF of the interevent times
 is obtained by means of Jacobi's theorem for univariate variable transformations as follows:
 \[f_{\tau}(\tau)= f_{\rm S}(\phi(\tau))\, | \phi'(\tau) |,\]
where $\phi'(\tau)$ is the derivative of the loading function.  In particular, if the crustal strength follows the
 \textit{Weibull distribution} with scale parameter $S_{s}$ and modulus $m_s$,  the ITD has a PDF that  is
 determined by the equation
 \begin{equation}
 \label{eq:ITD-general}
 f_{\tau}(\tau)= m_s\, | \phi'(\tau) | \, \frac{\phi(\tau)^{m_{s}-1}}{S_{s}^{m_{s}}} \, e^{- \left( \frac{\phi(\tau)}{S_{s}}\right)^{m_{s}}}.
 \end{equation}

 \subsection{Stress accumulation scenarios}

\subsubsection{Linear stress accumulation}

\begin{figure*}
\centering
\subfloat[]{\includegraphics[width=0.25\textwidth]{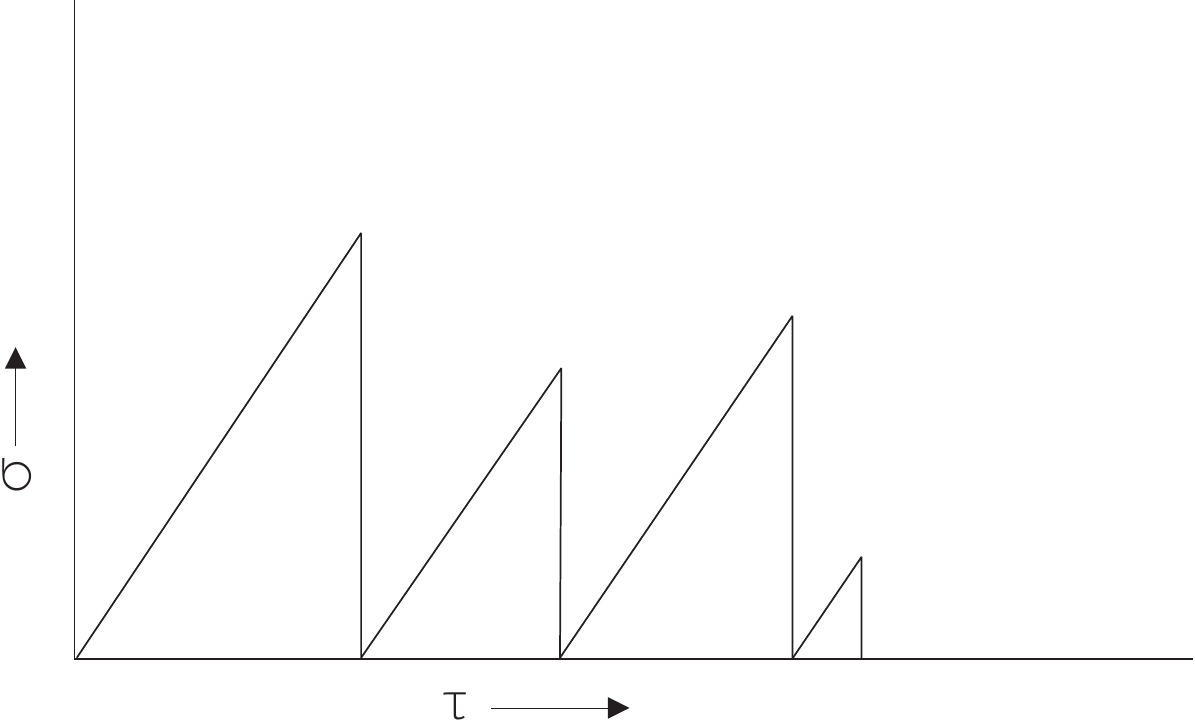}
\label{fig:accum-linear}}
\subfloat[]{\includegraphics[width=0.25\textwidth]{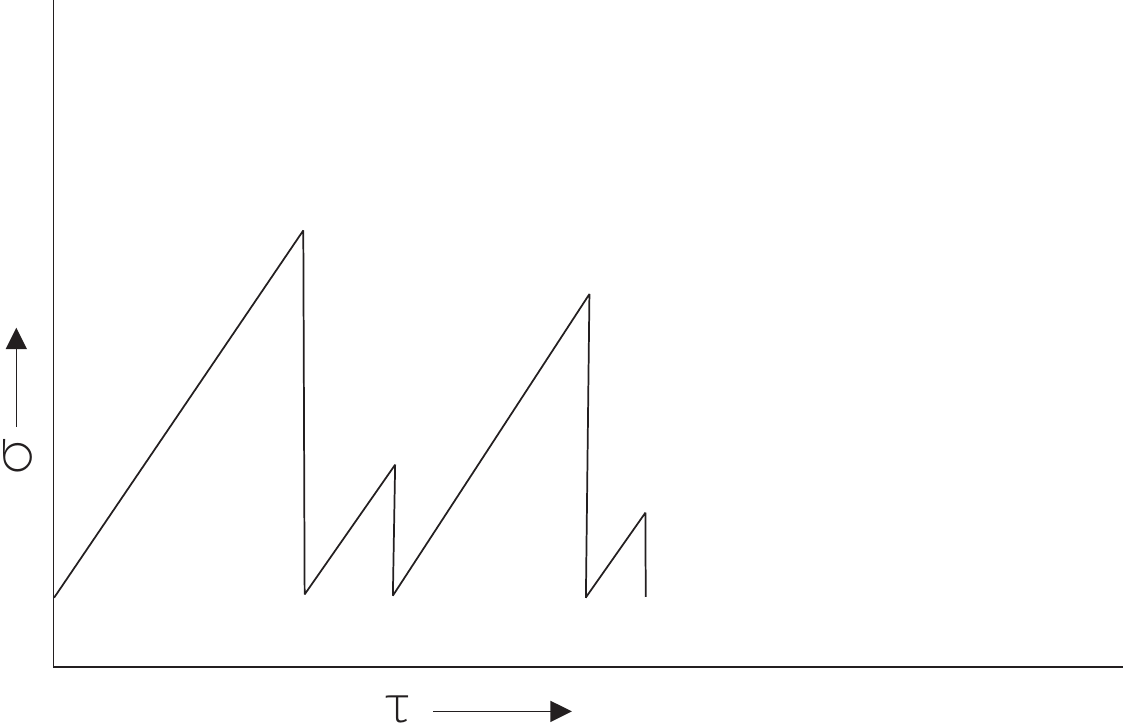}
\label{fig:accum-linear-res}}
\subfloat[]{\includegraphics[width=0.25\textwidth]{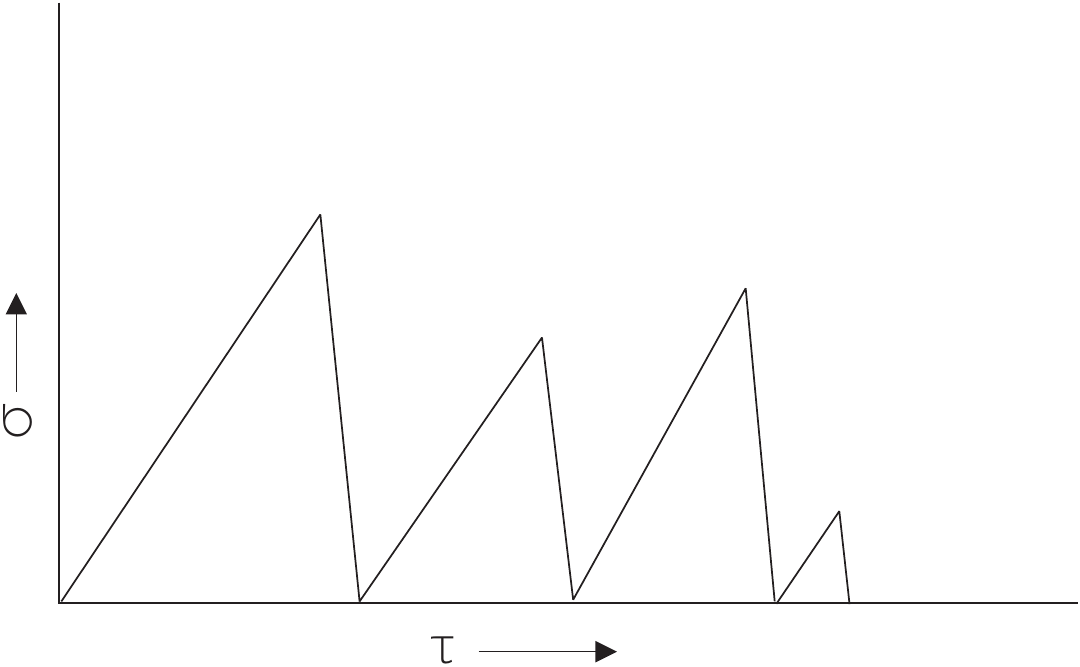}
\label{fig:accum-linear-rel}}\\
\subfloat[]{\includegraphics[width=0.25\textwidth]{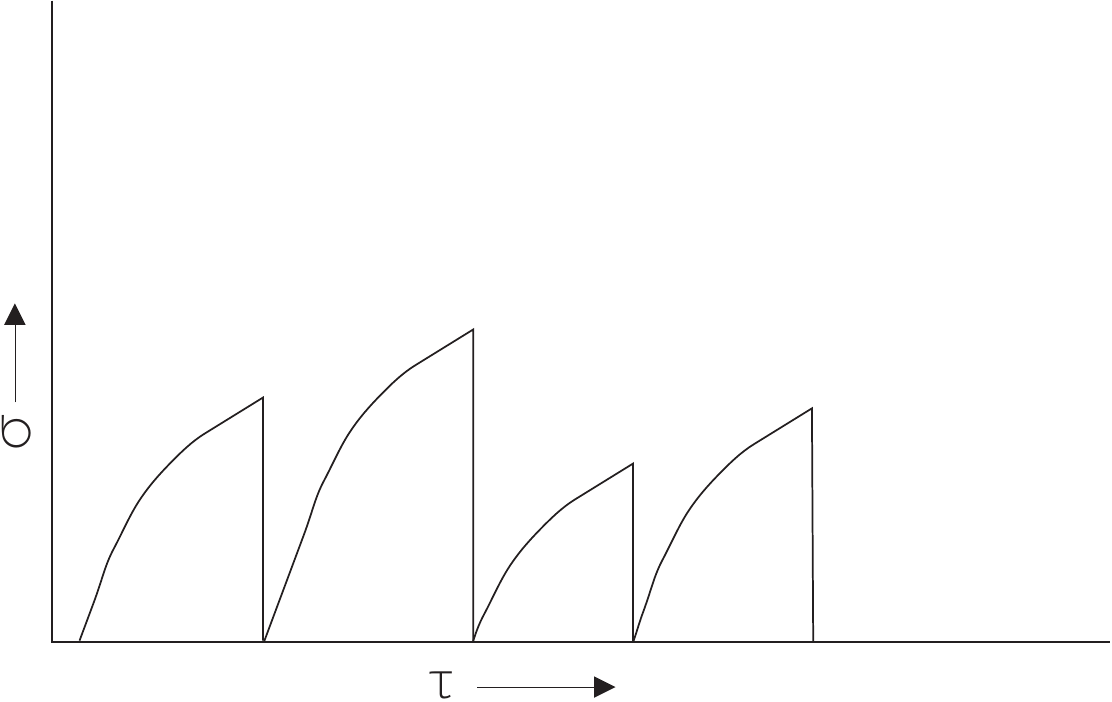}
\label{fig:accum-nonlinear}}
\subfloat[]{\includegraphics[width=0.25\textwidth]{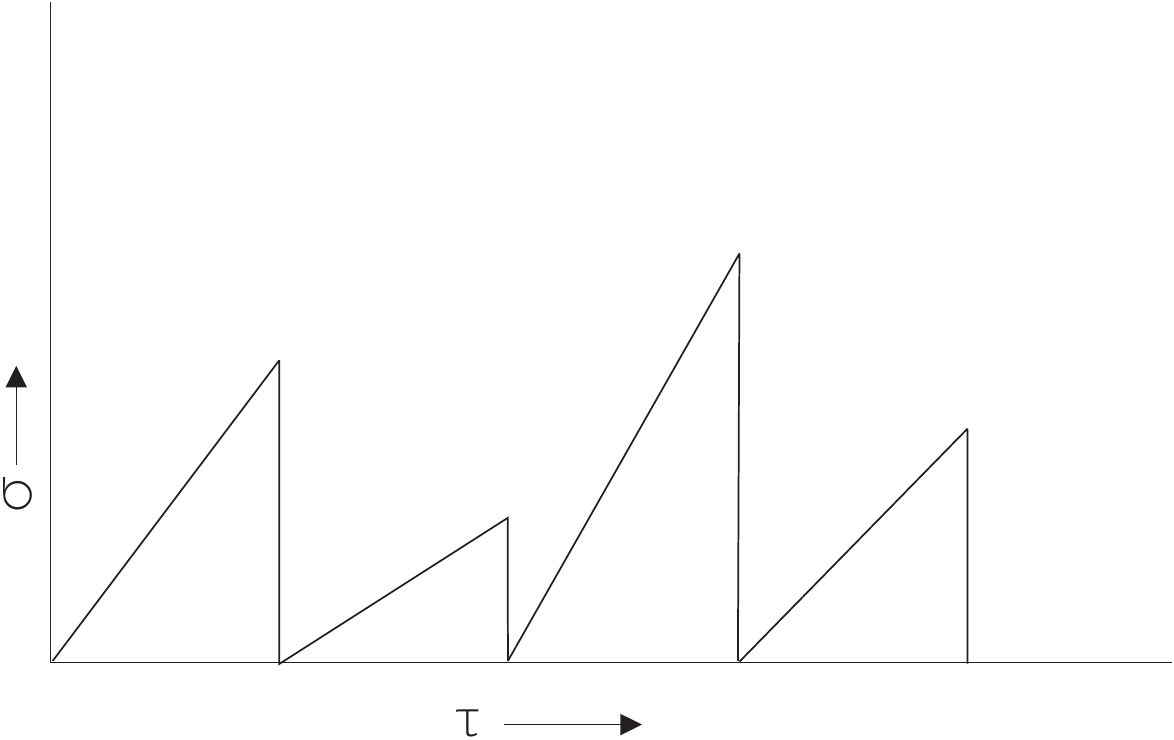}
\label{fig:accum-nonlinear-stochrate}}
\subfloat[]{\includegraphics[width=0.25\textwidth]{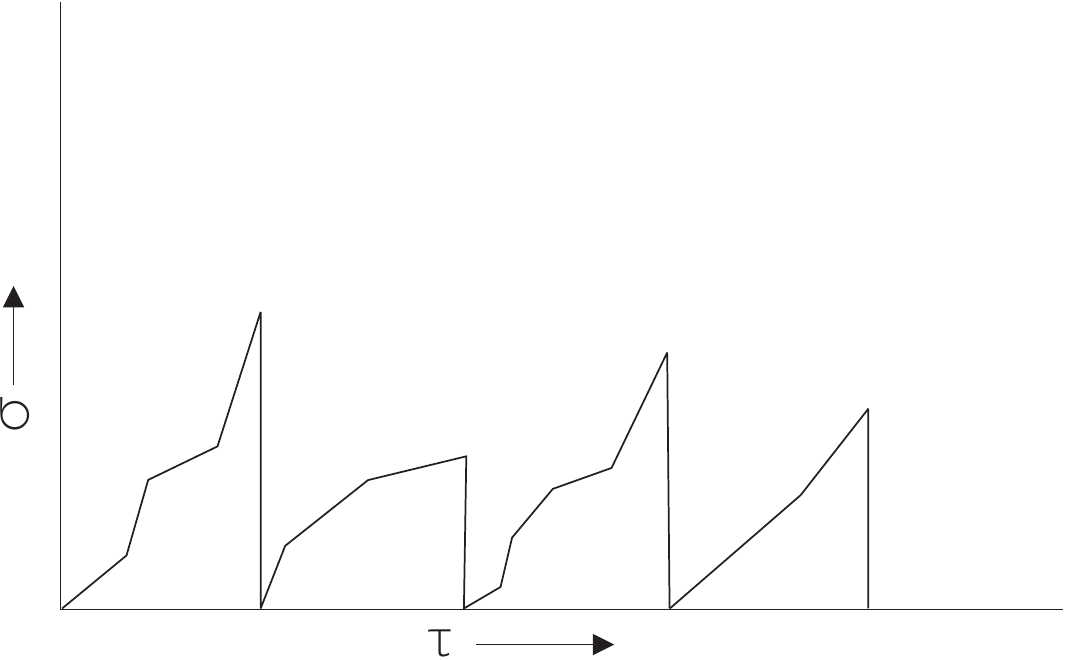}
\label{fig:accum-nonlinear-stocha}}
\caption{Schematic drawings of stress accumulation scenarios. (a) Linear loading function with zero residual stress and instantaneous relaxation. (b)  Linear loading function with finite residual stress
(c) Linear loading function with finite relaxation time. (d) Nonlinear loading function. (e) Linear loading function with stochastic rate. (f) Stochastic loading function. }
\label{fig:stress-accum}
\end{figure*}

\paragraph{Zero residual stress:} If $\phi(t) = \alpha \, \tau$, where $\alpha$ is the \textit{stress accumulation rate}, it follows from Eq.~\eqref{eq:ITD-general} that the ITD is the Weibull distribution with the CDF of Eq.~\eqref{eq:Weib-cdf}
with $\tau_s = S_s/\alpha$
and Weibull modulus $m=m_s$.  In this scenario the stress relaxes to zero after each seismic event,
c.f. Fig~\ref{fig:accum-linear}. Since the stress accumulation rate is independent of the strength, $\tau_s$ and $m$ may also vary independently.

 \paragraph{Finite residual stress:} If the stress relaxation process terminates at a non-zero residual stress,  $\sigma_{\rm res}$,
the loading function is  given by $\phi(t) = \sigma_{\rm res} + \alpha \, t$.
This scenario, illustrated in Fig.~\ref{fig:accum-linear-res}, is equivalent to the \emph{elastic rebound} theory developed by Reed~\cite{Reed10,Sornette99}; if a
fault with a shear modulus $G$ and width $2b$ is sheared at constant tectonic velocity $u_0$,  then
$\alpha = G\, u_0/ 2b$.  Since the residual stress does not  cause failure of the crust, the crustal strength is expected to have a minimum value $S_{\min} \ge \sigma_{\rm res}$. Hence, the strength distribution corresponds to a three-parameter Weibull, the CDF of which is given by Eq.~\eqref{eq:Weib-cdf}  with $\sigma$ replaced by $\sigma  - S_{\min}.$   In this case, based on Eq.~\eqref{eq:ITD-general} the ITD becomes
 \begin{equation}
 \label{eq:ITD-general-linear}
 f_{\tau}(\tau)= \frac{m}{\tau_s}\,   \ \left(  \frac{\tau-\tau_{\rm loc}}{\tau_s} \right)^{m-1}\, e^{- \left(  \frac{\tau-\tau_{\rm loc}}{\tau_s}  \right)^{m}} ,
 \end{equation}
where   $\tau_{\rm loc}= ( S_{\min} - \sigma_{\rm res} )/\alpha$
is the \emph{time location} parameter. If $S_{\min} = \sigma_{\rm res}$ the two-parameter Weibull model
is recovered. In addition, if $S_{\min} - \sigma_{\rm res}<< S_{s}$, the two-parameter Weibull is an accurate approximation. A value of $\tau_{\rm loc}$ that is not negligible compared to $\tau_s$ supports the use of the three-parameter Weibull ITD model in~\cite{Robinson09}.

\paragraph{Finite relaxation time:} The analysis above is not substantially modified if the relaxation has a finite duration $t^{\ast}_{i+1}$ (as shown schematically in Fig.~\ref{fig:accum-linear-rel}), and is governed by a decreasing function $\phi_{\rm rel}(t)$, since
 $t^{\ast}_{i+1}=(\phi_{\rm rel})^{-1}(S_{i+1})$ is also determined from the strength
 $S_{i+1}$. For example, assuming linear loading and relaxation relations, it follows that the stress at failure is given by
 $\sigma(\tilde{t}_{i+1}) = \alpha\,\tilde{t}_{i+1} $ and $\sigma(\tilde{t}_{i+1}) = \alpha_{\rm rel}\,\tilde{t}_{i+1}^\ast $. Hence, the total interevent time (measured between the zero stress values of consecutive phases) is given by $\tilde{t}_{i+1} + \tilde{t}_{i+1}^\ast,$ which is equivalent to replacing   $\alpha^{-1}$ with
$\alpha^{-1} + \alpha_{\rm rel}^{-1}$.  Hence,  in the following we simply use $\alpha$ without loss of generality.  In the case of aftershocks triggered by a main shock, the relaxation may involve a non-monotonic evolution of the stress toward its residual value. Then, $\alpha_{\rm rel}$ should be
considered as an effective relaxation rate and modeled as a  random variable.

\subsubsection{Nonlinear stress accumulation}

The following scenarios assume that the relaxation time is negligible and that the residual stress is zero.

\paragraph{Logarithmic stress accumulation:} This case is governed by a loading function $\phi(\tau) = \sigma^{\ast} \,\log(\tau/\tau_d),$ which implies sublinear stress evolution. According to Eq.~\eqref{eq:ITD-general},
logarithmic loading  leads to the \textit{log-Weibull} ITD
\begin{equation}
\label{eq:logW-ITD}
 f_{\tau}(\tau)= \frac{m_s}{\tau} \, \left( \frac{\sigma^{\ast}}{S_{s}} \right)^{m_{s}}\, \left[ \log\left(\frac{\tau}{\tau_d}\right) \right]^{m_{s}-1} \,
 e^{- \left( \frac{\sigma^{\ast} \,\log(\tau/\tau_d)}{S_{s}}\right)^{m_{s}}}.
\end{equation}

An equivalent expression has been investigated in~\cite{Hasumi09a,Hasumi09b}, where the authors employ a linear mixture of the Weibull and log-Weibull distributions to model the ITD for earthquakes in different tectonic settings.
As we have shown above, the log-Weibull is justified for a sub-linear stress accumulation.

\paragraph{Power-law stress accumulation:} This scenario, depicted in Fig.~\ref{fig:accum-nonlinear}, is governed by the loading function  $\phi(t) = \alpha\, \tau^{\beta},$ which leads to the \textit{Weibull} ITD

 \begin{equation}
\label{eq:W-ITD}
 f_{\tau}(\tau)= m \, \left( \frac{\tau}{\tau_{s}} \right)^{m-1} \,
 e^{- \left( \frac{\tau}{\tau_s}\right)^{m}},
\end{equation}
where $m = m_s\,\beta$ and  $\tau_s = (S_s/\alpha)^{1/\beta}.$ In the following, we do not distinguish between the linear and the power-law stress accumulations models since they both lead to a Weibull ITD.

Note that the relation $m = m_s\,\beta$ enables an indirect estimation of the loading exponent $\beta$
from the ITD Weibull modulus $m$, obtained from the analysis of the seismic event times, provided that the strength Weibull modulus is estimated from laboratory measurements.  For example, if $m_s =10$ and $m=0.7$, it follows that $\beta=0.07.$ Let us also use the transformation $\alpha = \lambda /\tau_{0}^{\beta}$, where $\lambda$ is a constant with units of stress and $\tau_{0}$ is an arbitrary time constant.  Then, it is possible to empirically determine $\lambda$. For example, if $S_s$ is known, e.g., $S_s = 250$ MPa (mega Pascal), $\tau_s = 1$day while $\tau_0$ is arbitrarily set to $\tau_0=1$sec,
it follows that $\lambda = S_s/(\tau_{s}/\tau_0)^{\beta} \approx 87.57 $ Pa.

 \paragraph{Stochastic stress accumulation:} In general, the statistical properties of the crust change over  geological time~\cite{Kanamori04}. Such changes on individual faults may be due to progressive
 damage caused by ongoing seismic activity~\cite{Serino11}. Hence, the parameters of the strength distribution may exhibit variations over time. In addition, the stress accumulation process may exhibit
 stochastic behavior over different time scales. For example, in the linear stress accumulation scenario the rate may fluctuate, as shown schematically in Fig.~\ref{fig:accum-nonlinear-stochrate}. More generally, the stress accumulation may have a complex dependence with intra-phase variations of the  accumulation rate as shown in Fig.~\ref{fig:accum-nonlinear-stocha}. In such cases, the fault is characterized by the effective ITD:
 \[
 F_{\tau}^{\ast}(\tau;{\bm \theta}^{\ast}) = \int_{{\bm \theta}_d}^{{\bm \theta}_u}  d{\bm \theta} \,
 p({\bm \theta}) \, F_{\tau}(\tau;{\bm \theta}),
 \]
 where ${\bm \theta}=(\theta_1, \ldots, \theta_k)$ is a parameter vector with joint PDF $ p({\bm \theta})$ that includes both strength and stress accumulation parameters. The function $ p({\bm \theta})$ is non-negative and integrable.
   If all the components of $ {\bm \theta}$ have a finite support and  $F_{\tau}(\tau;{\bm \theta})$ is a continuous function of ${\bm \theta}$, it is possible to iteratively apply the first mean-value theorem of integration, which leads to
 \begin{equation}
 \label{eq:renorm-CDF}
 F_{\tau}^{\ast}(\tau;{\bm \theta}^{\ast})= F_{\tau}\left(\tau; \theta_1^{\ast}(\tau), \theta_2^{\ast}(\tau,\theta_1^{\ast}), \ldots \theta_k^{\ast}(\tau,\theta_1^{\ast}, \ldots \theta_{k-1}^{\ast}) \right)
 \end{equation}
If the dependence of the effective parameters $\theta_{i}^{\ast}$ on $\tau$ is weak,
 the effective CDF is of the same functional form as $F_{\tau}(\tau;{\bm \theta})$, with ${\bm \theta}$ replaced by the \textit{effective parameter vector} ${\bm \theta}^\ast$ whose components are inter-dependent.

\subsection{Magnitude dependence of ITD on single faults}
In the following, we refer to the interevent times for the complete (i.e., including all magnitudes)
seismic sequence as \textit{primitive interevent times}.
The main assumption in the analysis is that the primitive ITD conforms to the Weibull distribution.

\subsubsection{Impact of magnitude threshold on sampled strength distribution}
To link the model of repeated stress accumulation and relaxation phases with the ITD, we have
 assumed that all the  seismic events are included, regardless of their magnitude.
Then, the interevent times sample  the entire crust strength distribution.  In practice (see Section~\ref{sec:Crete}),  one focuses on seismic events over a \textit{threshold magnitude} that varies depending on the area, the resolution of the instrumental network, and the goals of seismic risk assessment. Often the threshold magnitude is chosen as the \textit{magnitude of completeness} above which all events are resolved by the observation system~\cite{Woessner05}.

The stress drop caused by an earthquake  is  related to the earthquake magnitude by  means of empirical,  monotonic relations~\cite{Kanamori75}. Then, assuming that $\sigma_{\rm res}$ is constant,
 the threshold magnitude, $M_{L,c}$, corresponds
 to a unique crustal strength value $S_c$. Hence, the proportion of earthquakes
 exceeding in magnitude $M_{L,c}$ is  $1 - F^{\ast}_{\rm S}(S_c)$.
The sampled PDF for events with $S > S_c$ is given by the following equation
\begin{equation}
\label{eq:mod-weib}
f^{\ast}_{\rm S > S_c}(S)=\frac{f^{\ast}_{\rm S}(S)}{1-F^{\ast}_{\rm S}(S_c)} \, \vartheta(S - S_c),
\end{equation}
 where $\vartheta(\cdot)$ is the unit step function and the denominator normalizes the PDF.
 The  removal of  low strength values implies that the respective $\Phi(S)$ has a concave lower tail, even if the effective fault distribution $F^{\ast}_{\rm S}(S)$ is Weibull. In this respect, the impact of coarse-graining by means of a finite threshold is similar to that of non-resolved, low-magnitude events.

\begin{figure}
\includegraphics[width=0.65\textwidth]{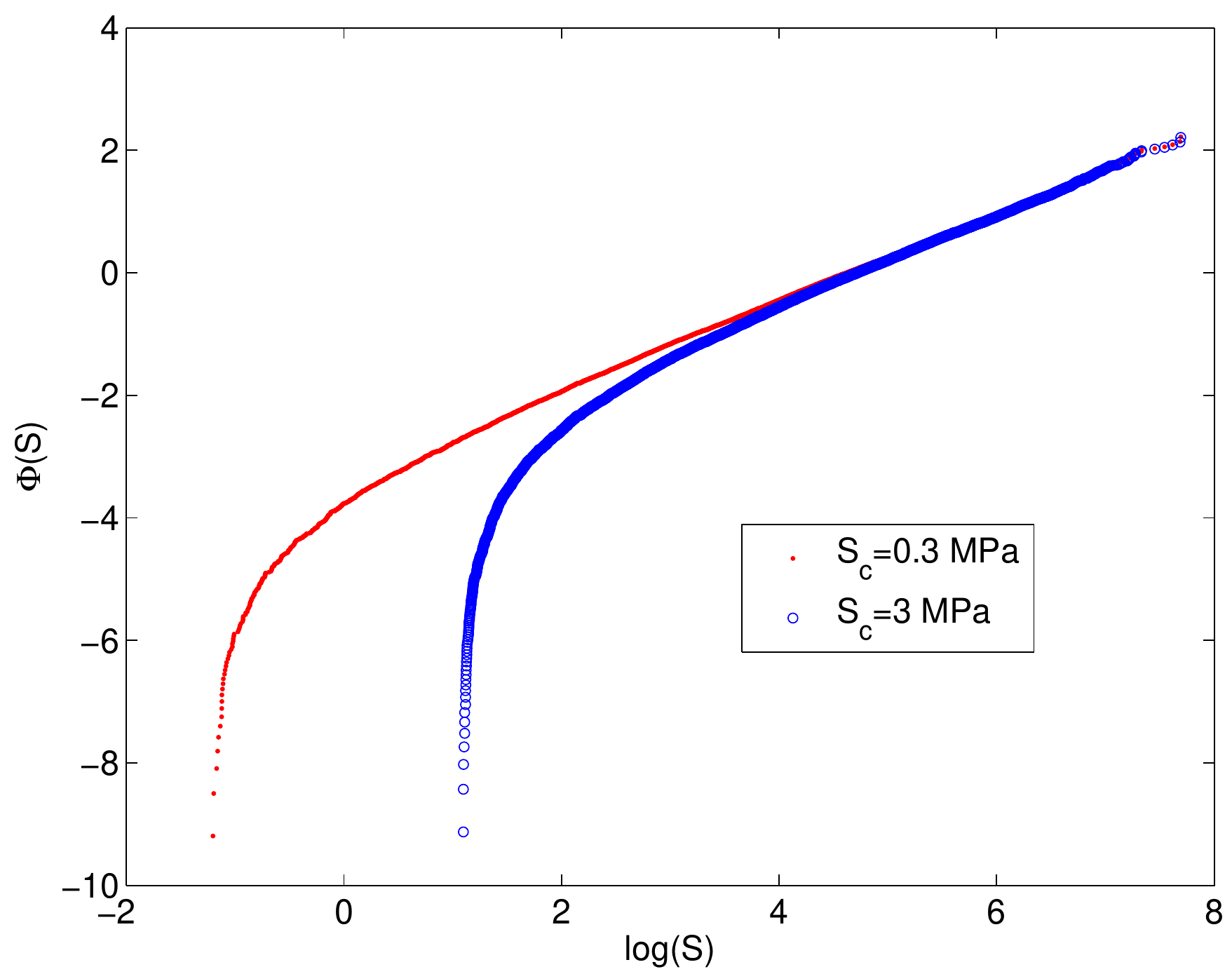}
\caption{\label{fig:turn_wbl} Plots of $\Phi(S)$ derived from a sample of 10.000 Weibull random numbers
from a  distribution with $S_s= 100$ MPa and $m_s=0.7$ using two different thresholds ($S_{c} = 300$ KPa, $3$ MPa). }
\end{figure}

In Fig.~\ref{fig:turn_wbl} we show the impact of coarse-graining on a sample of Weibull random numbers with $S_s= 100$ MPa (which is a typical average value of normal-fault strength~\cite{Sibson74}) and $m_s=0.7$ with cutoffs  $S_{c} = 300$ KPa and $3$ MPa. In spite of the curvature of $\Phi(S)$ below the threshold, the linear part of both $\Phi(S)$  plots has the same slope, i.e., the same Weibull modulus. Nevertheless, the estimated modulus~\footnote{The Matlab maximum likelihood estimator is used} based on the truncated sequence increases with $S_{c}$; namely, $\hat{m}_s \approx 0.74$ for $S_c = 0.3$~MPa and $\hat{m}_s \approx 0.83$ for $S_c = 3$ MPa. Hence, an increasing $S_c$ leads to an  increase of the estimated Weibull strength modulus.

\subsubsection{Impact of magnitude threshold on ITD}

The times between events exceeding $M_{L,c}$  extend to several stress accumulation and relaxation cycles.
We assume a stress accumulation phase between events with $M_{L} \ge M_{L,c}$, such that stress increases non-monotonically between two events from a low residual value to the strength threshold that corresponds to
$M_{L}$. Then, we can replace the non-monotonic dependence with an ``effective'' monotonically increasing stress accumulation function; e.g., in the linear case, $\alpha$ is replaced by $\alpha_{\rm eff}$. Consequently, the ITD for finite $M_{L,c}$ is also determined from Eq.~\eqref{eq:mod-weib}. In general,  $\alpha_{\rm eff}$ should be treated as a random variable that fluctuates depending on the number of sub-threshold events contained between the two supra-threshold events.  A power-law stress accumulation function transfers to the ITD the lower-tail curvature of the crustal strength resulting from the truncation of above-threshold events.

The interevent times between events exceeding $M_{L,c}$ can be treated as sums of a randomly varying number of primitive times. To our knowledge, very little is known about sums of Weibull random numbers: a review of known results for sums of random variables reports the absence of  such results for the Weibull distribution~\cite{Nadarajah08}. Recently,   convergent series expansions for the PDF that governs the sum of Weibull random numbers were obtained in~\cite{Yilmaz09}.  The situation is more complicated for earthquake interevent times, since the number of summands, which is determined by the excursions of the dynamic stress above $S_c$,  fluctuates. This coarse-graining operation completely eliminates strength values below $S_c$,  in contrast with the summation of a fixed number of variables that shifts the mode of the distribution but does not eliminate the weight of the PDF near zero.

 \section{ITD of fault systems}
 \label{sec:ITD-system}

Above, we  focused on the one-body ITD problem by concentrating on single faults.
 Nevertheless, the seismic behavior of a given area is a many-body problem that involves multiple faults.
 The interevent times of  a fault system  are not directly obtained from the interevent times of the individual faults in the system. Let us assume that a fault $Fa$  hosts two seismic events at times $t^{Fa}_{j}$ and $t^{Fa}_{j+1}$ leading to an interevent time $\tau^{Fa}_{j}=t^{Fa}_{j+1}- t^{Fa}_{j}.$
 In addition, let fault $Fb$ host two seismic events at times $t^{Fb}_{i}$ and $t^{Fb}_{i+1}$,
 with an interevent time $\tau^{Fb}_{i}=t^{Fb}_{i+1}- t^{Fb}_{i}.$ Next, let us assume that
 $t^{Fa}_{j} <  t^{Fb}_{i} < t^{Fb}_{i+1} < t^{Fa}_{j+1}.$ Then, the respective
 interevent times for the two-fault system are
 $t^{Fb}_{i} - t^{Fa}_{j}, \; t^{Fb}_{i+1} - t^{Fb}_{i}, \; t^{Fa}_{j+1} - t^{Fb}_{i}$, which can not be obtained from
 $\tau^{Fa}_{j}$ and $\tau^{Fb}_{i}$.

 Nevertheless, we can extend the single-fault interevent-times expressions  to fault systems by assuming that all the faults are subject to a \emph{uniform stress accumulation process}. Then, the same loading function applies to the entire system, and the ITD is determined from the \textit{composite strength distribution} of the system~\footnote{This assumption is not valid if the stress accumulation is driven by fluid diffusion in the fault system}. Assuming that the system involves $N$ faults
 characterized by $K$ different effective  strength distributions
 $F^{\ast}_{\rm S}(S;{\bm \theta}_i), \; i=1,\ldots,K$ and that $M_i$  is the number of faults that follow the
 probability distribution $F^{\ast}_{\rm S}(S;{\bm \theta}_i)$ (i.e., $M_1 + M_2 + \ldots M_K=N$), the composite strength distribution of the fault system is given by the following superposition
 \begin{equation}
 \label{eq:CDF-fault-system}
 {\mathcal F}^{\ast}_{\rm S}(S) = \sum_{i=1}^{K} p_{i} \, F^{\ast}_{\rm S}(S;{\bm \theta}_i),
 \quad p_{i}= \frac{M_{i}}{N}, \, i=1, \ldots K.
 \end{equation}
 A \textit{homogeneous fault system} comprises  faults that share the same strength distribution, $F^{\ast}_{\rm S}(S;{\bm \theta})$. Then, the composite strength distribution is given by ${\mathcal F}^{\ast}_{\rm S}(S)=F^{\ast}_{\rm S}(S;{\bm \theta})$.

\subsection{System with bimodal strength distribution}
 Let us consider a system that involves faults governed mainly by two Weibull strength distributions with different parameters, so that $K_1$ faults follow the first distribution, while $K_2 = N - K_1$ faults follow the second. The composite strength distribution of the system is a bimodal Weibull.
 Let us assume that the survival function is given by the bimodal expression
 \begin{equation}
 \label{eq:CDF-fault-system-bimodal}
1-  {\mathcal F}^{\ast}_{\rm S}(S) = p \, e^{-\left(  \frac{S}{S_{1}}\right)^{m_{1}}}
+(1- p) \, e^{-\left(  \frac{S}{S_{2}}\right)^{m_{2}}},
 \end{equation}
where $p=K_{1}/M$.  This mixture model can lead to the ``M-smile'' histogram sometimes observed in the analysis of
 earthquake interevent times, e.g.~\cite{Naylor09}.

\begin{figure}
\centering
\begin{overpic}[width=0.75\textwidth]{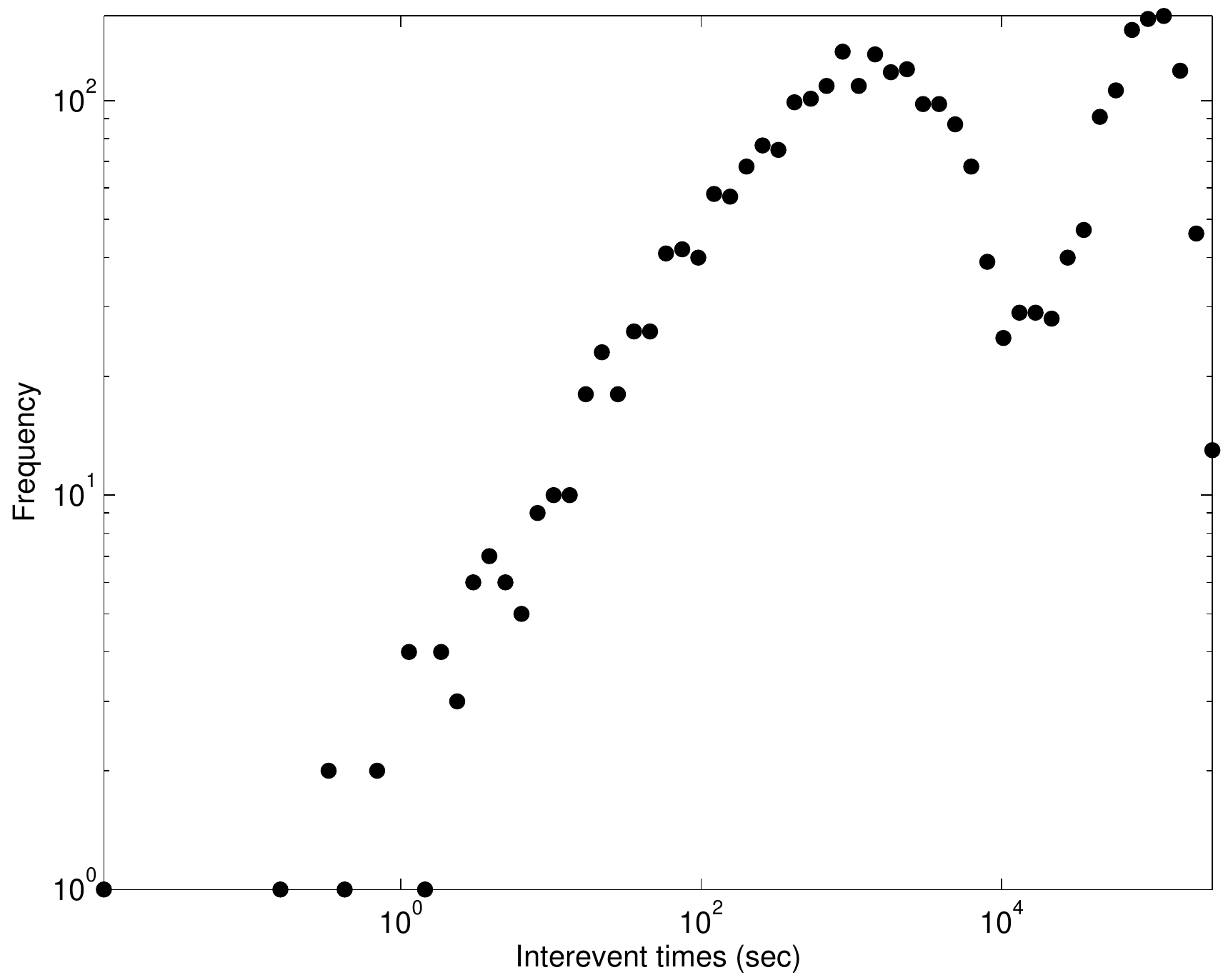}
\put(52,10){\includegraphics[scale=.2]%
{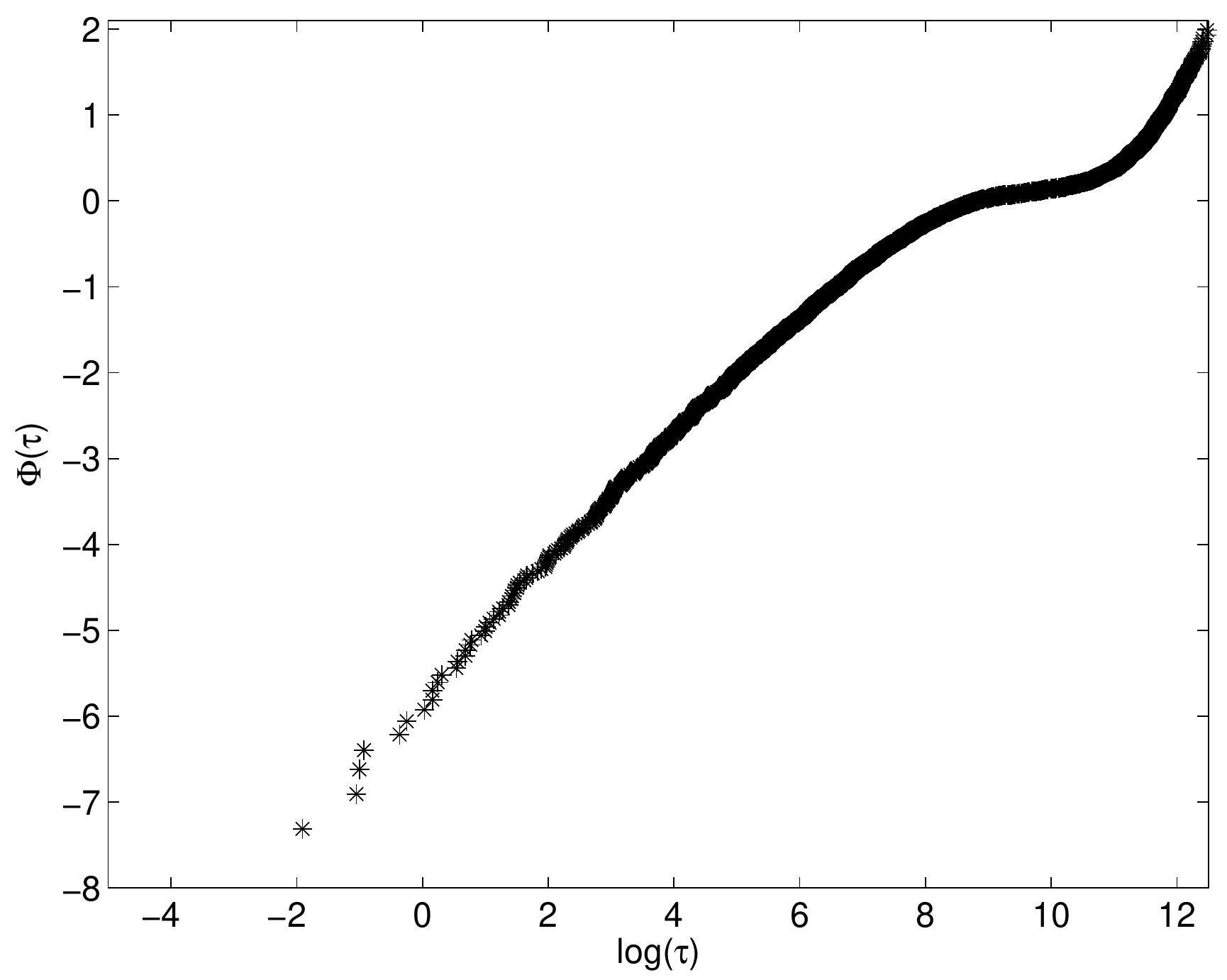}}
\end{overpic}
\caption{\label{fig:bimodal} Frequency histogram (log-log plot) and Weibull plot (inset) for $3000$ random numbers simulated from a  bimodal Weibull  mixture model (see text for parameters). }
\end{figure}

 In Fig.~\ref{fig:bimodal}  we show the histogram and corresponding
 empirical $\hat{\Phi}(\tau)$ for a set of $3 000$ interevent times generated from the bimodal Weibull distribution
 with $p=2/3,$ $m_1=0.7$, $m_2=1.8$, $\tau_{1}=1.5\, 10^3$ (sec), and  $\tau_{2}= 10^5$ (sec).
The signature of bimodality is apparent as a dip in the histogram and as a saddle point in the Weibull plot.
The parameters of a bimodal ITD can be estimated using the expectation-maximization algorithm~\cite{Dempster77}.

The concept of a bimodal ITD has been proposed in~\cite{Naylor09}, where it is invoked to represent the mixing of correlated events (aftershocks) and uncorrelated (main)  events. In our model, there is no
\emph{a priori} distinction between aftershocks and main events. However, it is conceivable that main
events lead to temporary changes of the stress accumulation or the crustal strength, which imply a distinct
component  in the ITD linked with aftershock activity.

\subsection{Magnitude dependence of ITD for fault systems}
\label{ssec:mag-dep}

For a single fault we showed above that the lower tail of the sampled strength PDF is  cut-off
(c.f. Fig~\ref{fig:turn_wbl}).
 We expect that the abrupt  change in the slope of  $\Phi(S)$ at $S \approx S_c$ is reduced in seismic data from fault systems due to non-homogeneity of the strength parameters and possible non-stationarity caused by fluctuations in the Weibull parameters over time.

In the following, we assume that the fault-system ITD for events exceeding
 $M_{L,c}$  is approximated by the Weibull. The total duration of the complete seismic sequence
 is $T=\sum_{i=1}^{N} \tau_{i}$.
Let $\overline{\tau_1}$ and $\overline{\tau_2}$ denote sample average times
 corresponding to truncated seismic sequences obtained for  ${M}_{L,1} <  {M}_{L,2},$ respectively.  We assume
 $\overline{\tau_1}$ and $\overline{\tau_2}$ to be   accurate estimates of the ensemble means of each sequence, i.e.,
$\overline{\tau_1}/\overline{\tau_2}\approx \E[\tau_1]/\E[\tau_2]$.
If $N_{i}$ is the number of events with magnitude above ${M}_{L,i}$,
then $\overline{\tau_i} \approx T/N_{i}$.
We assume that the Gutenberg - Richter scaling,  $N_{1}/N_{2} = 10^{-b(M_{L,1} - M_{L,2})},$ is valid for the system. Then, it follows that
\begin{equation}
\label{eq:tau-mean-scaling}
\frac{\overline{\tau_2}}{\overline{\tau_1}}  \approx   \frac{\E[\tau_2]}{\E[\tau_1]}= e^{- \rho_M \, (M_{L,1} - M_{L,2})},
\end{equation}
where $\rho_M = b\, \log10$. Since for seismically active regions $b \in [0.5 - 1.5]$, it follows that
$\rho_M \in [1.15 - 3.45].$

The mean value of the Weibull distribution is given by $\E[\tau] = \tau_{s} \, \Gamma(1 + 1/m).$
Hence,
$\tau_{s,2}/\tau_{s,1}= \E[\tau_2]\, \Gamma(1 + 1/m_1)/\E[\tau_1]\, \Gamma(1 + 1/m_2).$ Then, based on Eq.~\eqref{eq:tau-mean-scaling} it follows  that
\begin{equation}
\label{eq:taus-scaling}
\frac{\tau_{s,2}}{\tau_{s,1}} \approx \frac{\Gamma(1 + 1/m_1)}{\Gamma(1 + 1/m_2)} \, e^{- \rho_M \, (M_{L,1} - M_{L,2})}.
\end{equation}
Equation~\eqref{eq:taus-scaling} also holds for single faults
that obey Gutenberg - Richter scaling and have a Weibull strength distribution.

\section{Analysis of Cretan Micro-Earthquake Sequences}
\label{sec:Crete}

\begin{figure}
\includegraphics[width=0.70\textwidth]{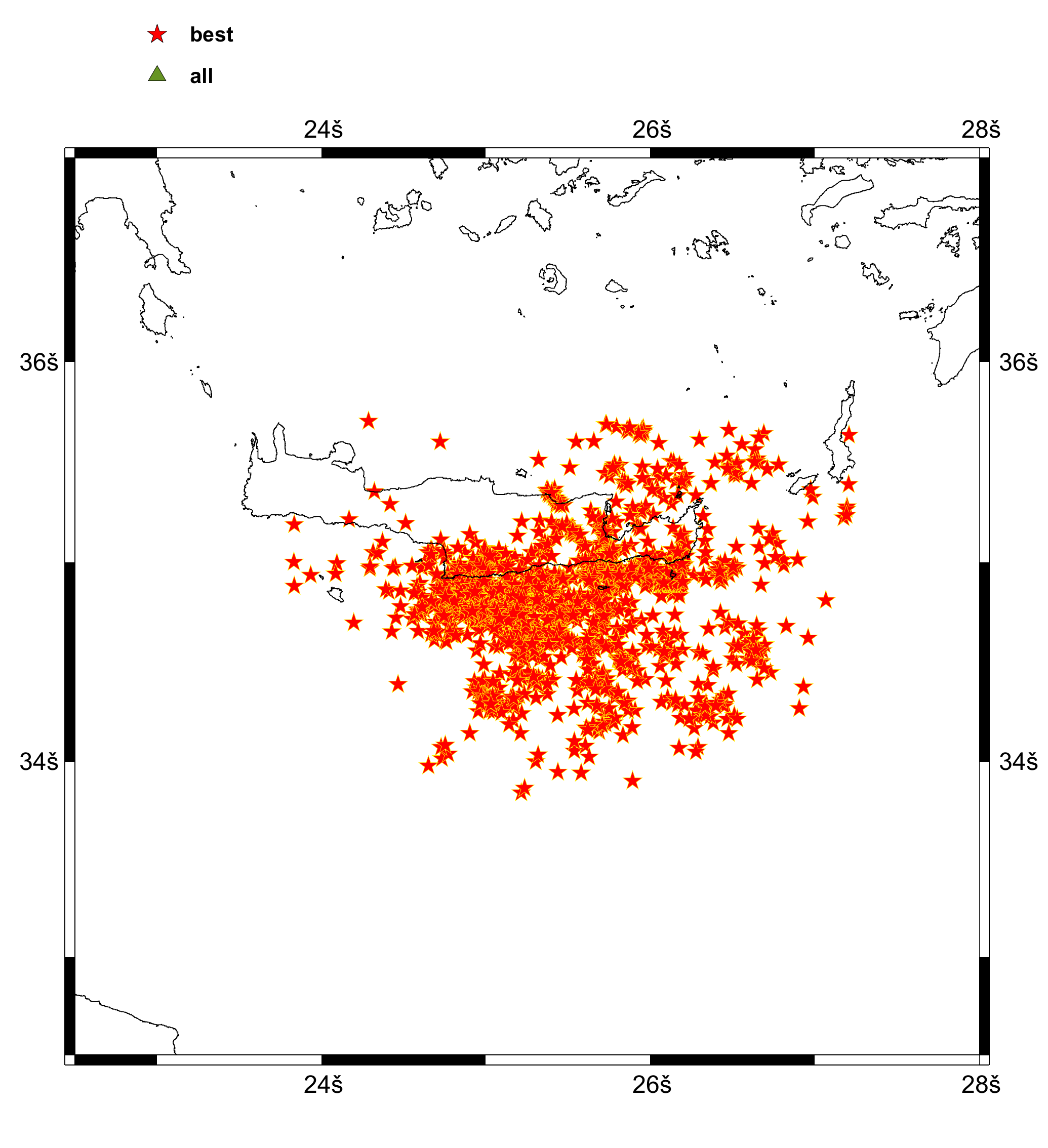}
\caption{\label{fig:kriti_locs} Map locations of CSS events (stars) on and around the island of
Crete. }
\end{figure}

The seismic data in the \textit{Cretan seismic sequence (CSS)} investigated below are from Becker et al.~\cite{Becker10}. The CSS resulted from tectonic activity generated at the Hellenic subduction margin, where the African plate is being subducted beneath the Eurasian plate; this is the seismically most active region in Europe.  More than 2\,500 local and regional \textit{micro-earthquake} events with magnitudes up to  4.5 ${ M_L}$ (Richter local magnitude scale) occurred during the time period between July 2003 and June 2004.
The micro-earthquakes were accurately recorded by an amphibian seismic network zone onshore and offshore Crete. The network configuration consisted of up to eight ocean bottom seismometers as well as five temporary short-period and six permanent broadband stations on Crete and smaller surrounding islands (e.g. Gavdos). The magnitude of completeness varies between 1.5 $M_L$ on Crete and adjacent areas and  2.5 $M_L$ at around 100 km south of Crete. Most of the seismic activity is located offshore of central and eastern Crete  (see Fig.~\ref{fig:kriti_locs}).
The repeat times between successive earthquake events range from 1 sec
to 19.5 days.

\subsection{Exploratory analysis}
\begin{figure*}
\centering
\subfloat[CSS interevent times]{\label{fig:Crete_times}\includegraphics[width=0.47\textwidth]{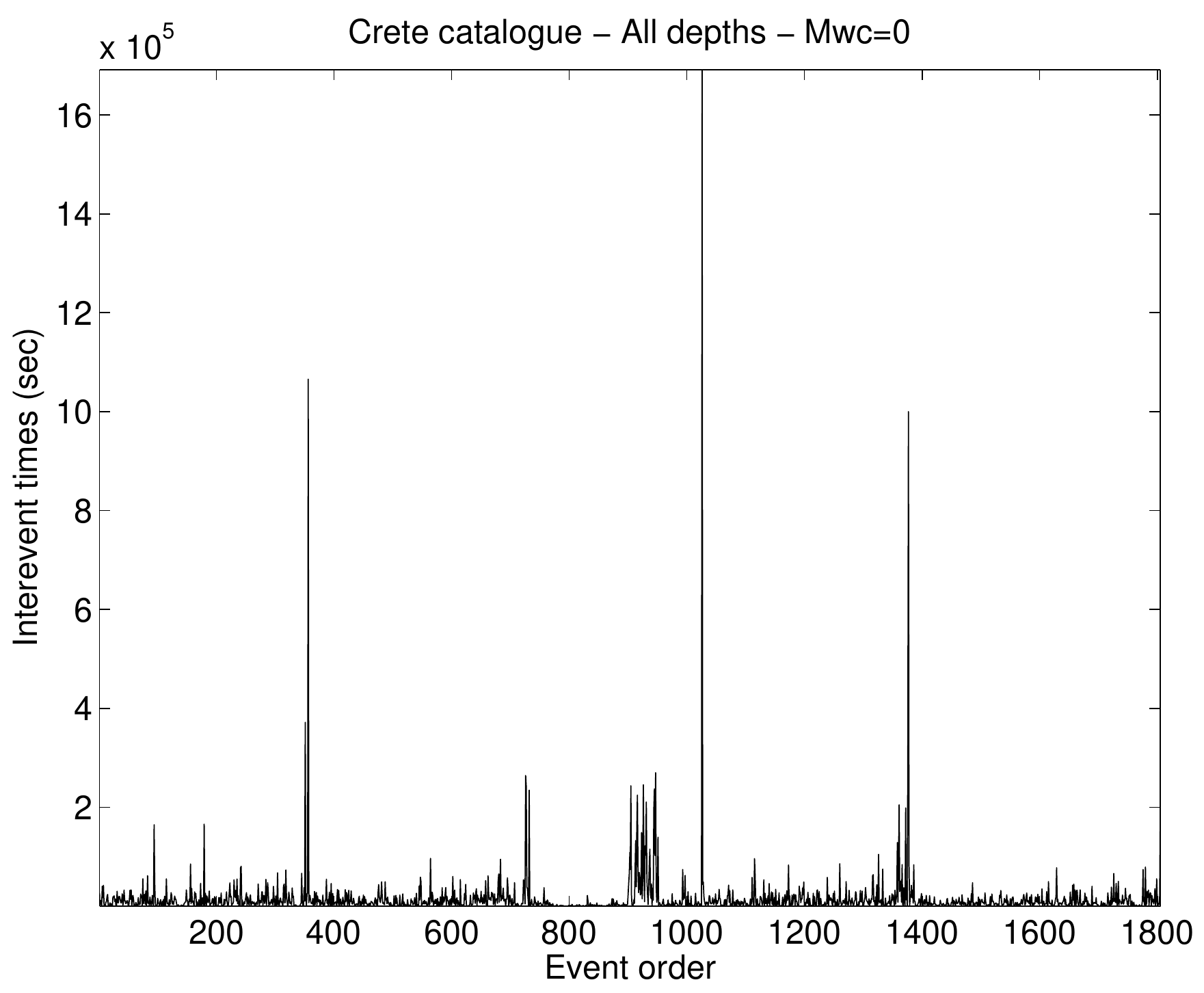}}
\subfloat[CSS focal depth]{\label{fig:depth}\includegraphics[width=0.47\textwidth]{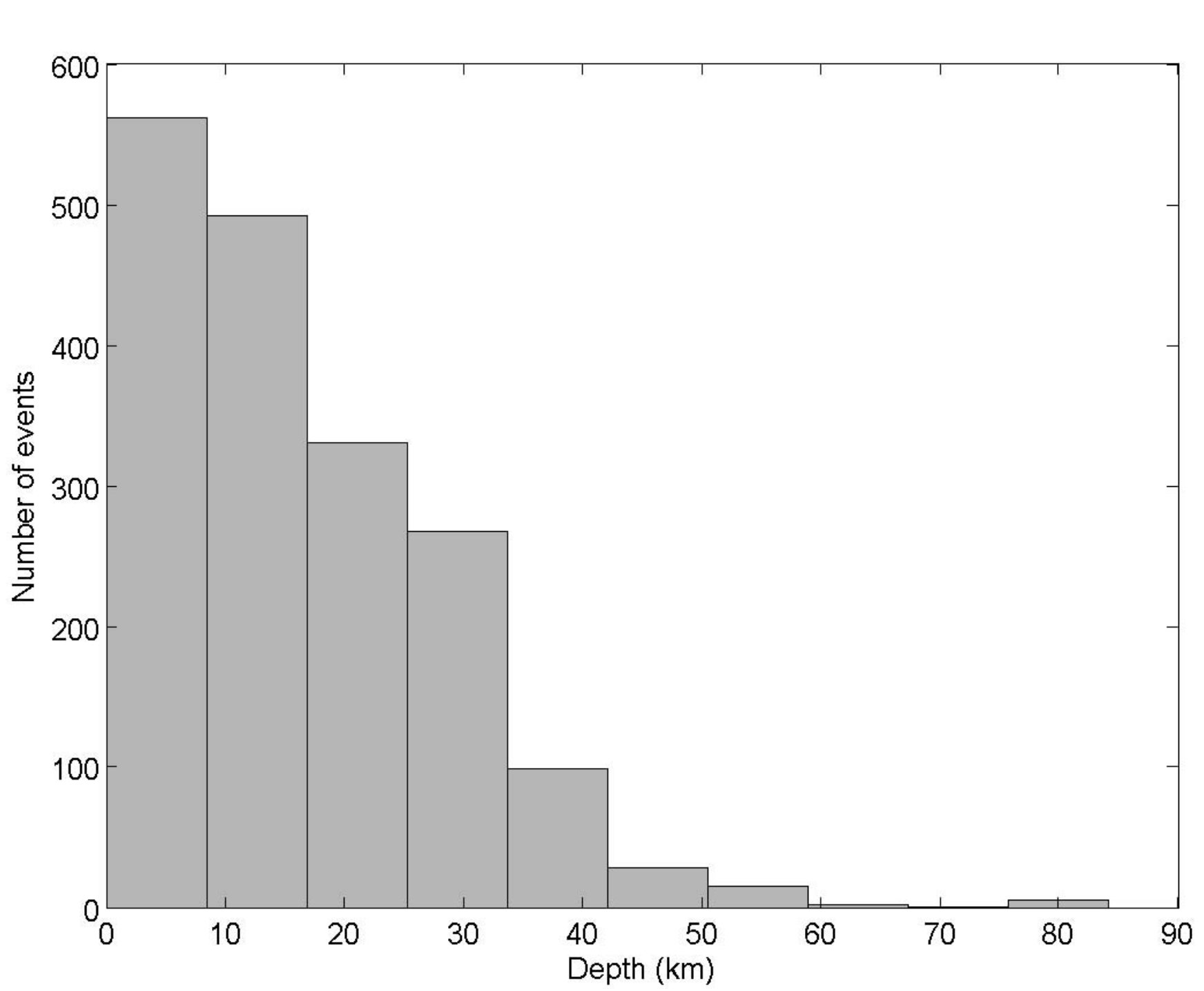}}
\caption{(a) Sequence of interevent times: the horizontal axis counts the order $(i=1, \ldots,N)$ of
the interval $\tau_{i}=t_{i+1}- t_{i}$, and the vertical axis measures $\tau_{i}.$ (b)
Histogram of the focal depths of earthquake events. }
\label{fig:Crete_data}
\end{figure*}

The exploratory analysis of CSS includes all the events in
the catalogue (magnitudes $>1 M_L$). There is no discernible main shock in the sequence, hence declustering of the
data is not considered.
We use the concept of natural time to measure the ordering of the event times~\cite{Varotsos05b,Varotsos05c}:
if $t_{k}$ is the time of the $k$-th event, the \textit{interevent time series} in natural time
is given by $\tau_{k}:=t_{k+1}-t_{k}$.
 The CSS  interevent times series is shown in Fig.~\ref{fig:Crete_times},
which exhibits isolated large peaks separating clusters of
almost continuous seismic activity. The distribution of the earthquake focal depths is shown in Fig.~\ref{fig:depth}.
Assuming a uniform tectonic stress over depth, the observed declining trend~\footnote{Depending on the bin size selected, the number of events may not decrease monotonically,
but the overall declining trend persists} agrees qualitatively with the reported increase of the crust strength
with depth~\cite{Sibson74,Zoback93}.

\begin{figure*}
\centering
\subfloat[]{\label{fig:Crete_Hist_Mw_all}\includegraphics[width=0.45\textwidth]{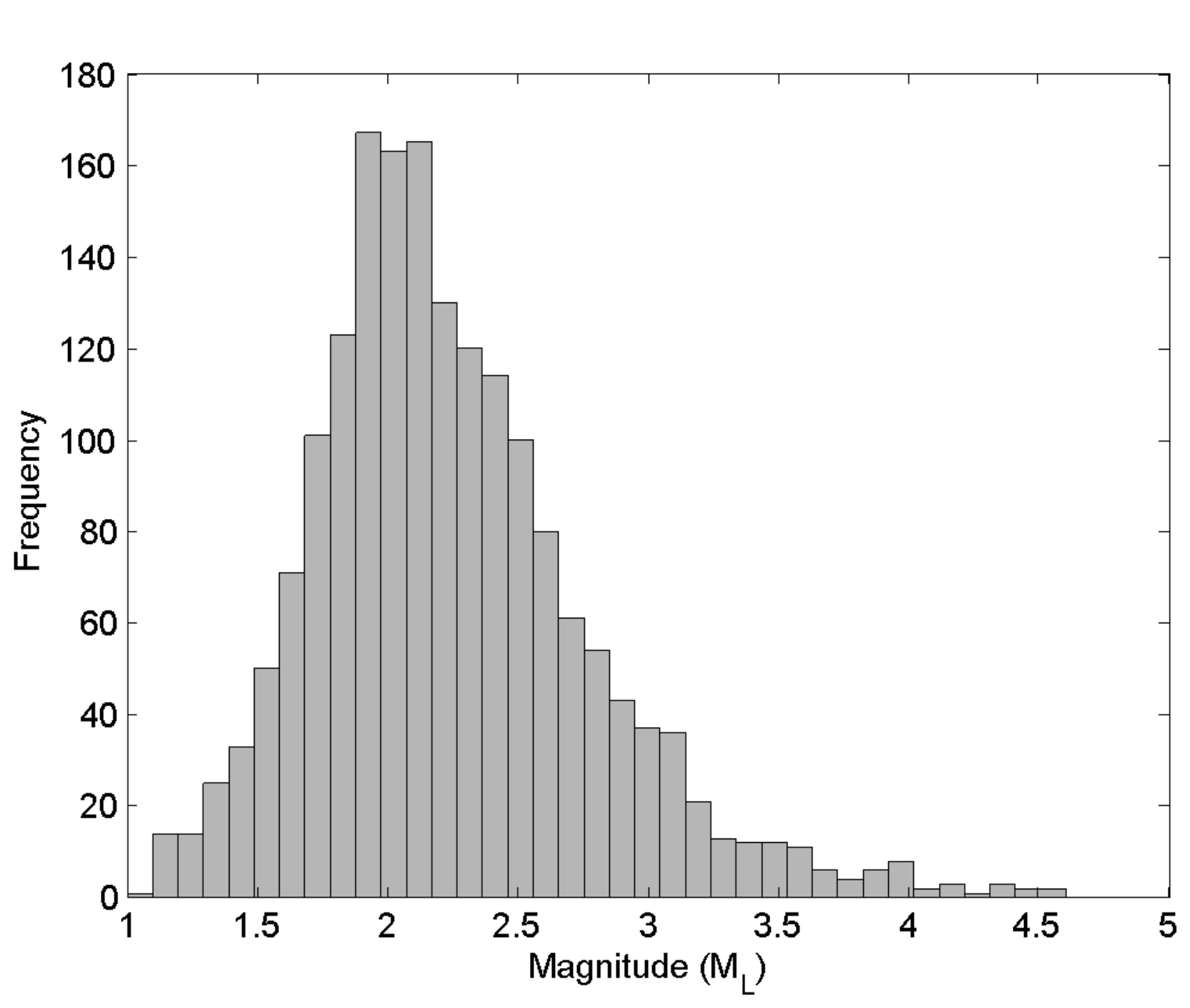}}
\subfloat[]{\label{fig:Crete_GR}\includegraphics[width=0.45\textwidth]{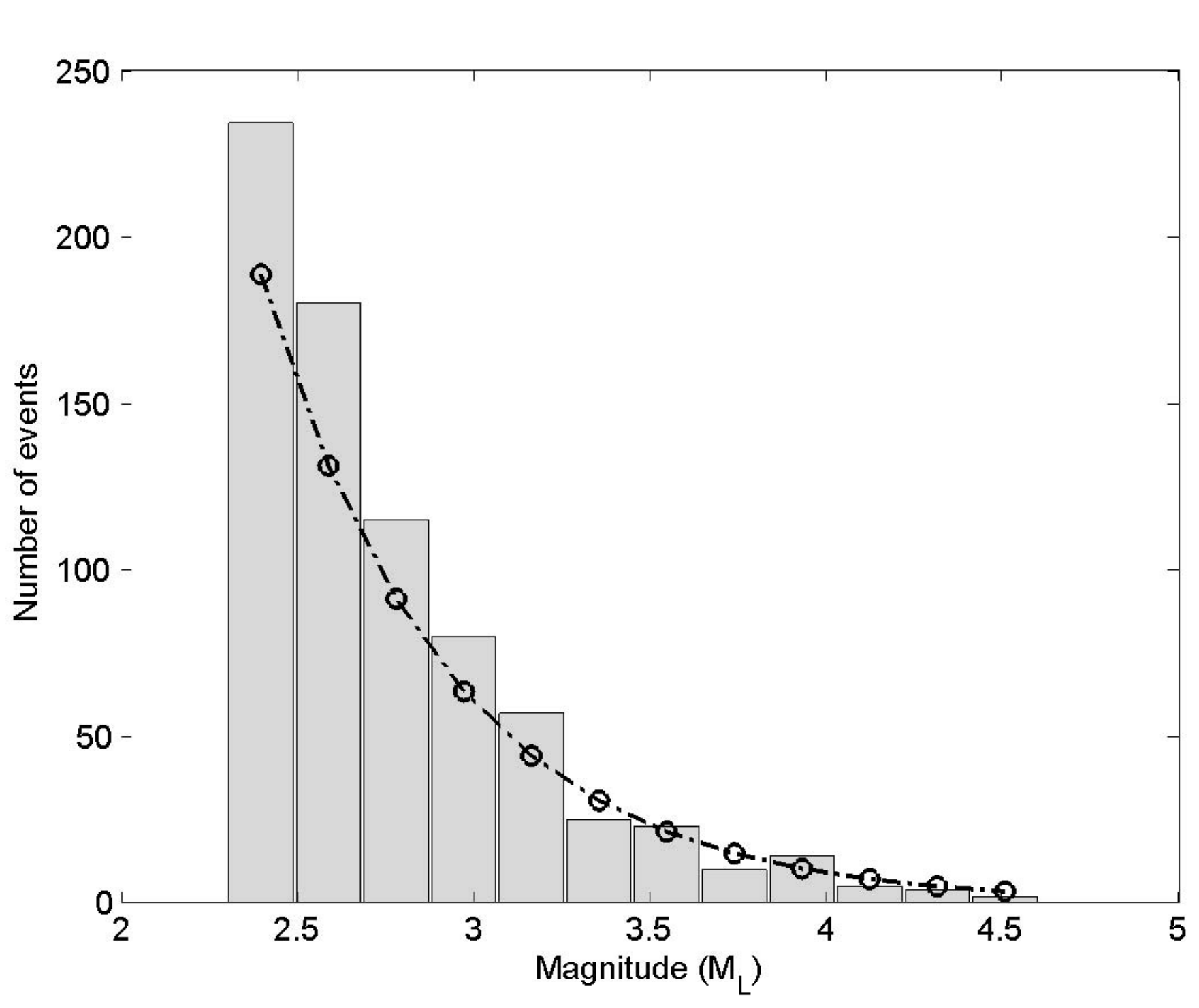}}
\caption{(a) Frequency histogram of events (complete CSS) versus magnitude. (b)
Histogram of event number versus threshold magnitude and best-fit line to Gutenberg - Richter model. }
\label{fig:magnitude}
\end{figure*}

The histogram of the magnitudes is shown in Fig.~\ref{fig:Crete_Hist_Mw_all}. The
Gutenberg - Richter law is expressed as $\log_{10}N({M_L}> M) = a - b\, M,$ where $N$ is the number of events with magnitude greater or equal to $M$. This implies an exponential decrease of $N$ with the magnitude, while  Fig.~\ref{fig:Crete_Hist_Mw_all} shows a non-monotonic dependence of $N$ on $M$.
This is due to the \textit{resolution problem}, namely the inability to  observe all events with magnitudes below a critical threshold that is typically around 2 $M_L$ and  $\approx 2.2 M_L$ for the CSS.
Setting the critical threshold at 2.2 $M_L$, the histogram of event frequency versus magnitude is shown in  Fig.~\ref{fig:Crete_GR}.
The maximum likelihood estimate of the Gutenberg - Richter exponent is $\hat{b} \approx 0.85$, i.e.,
 within the typical range $[0.5 - 1.5]$ for seismically active regions.

\begin{figure}
\centering
\includegraphics[width=0.75\textwidth]{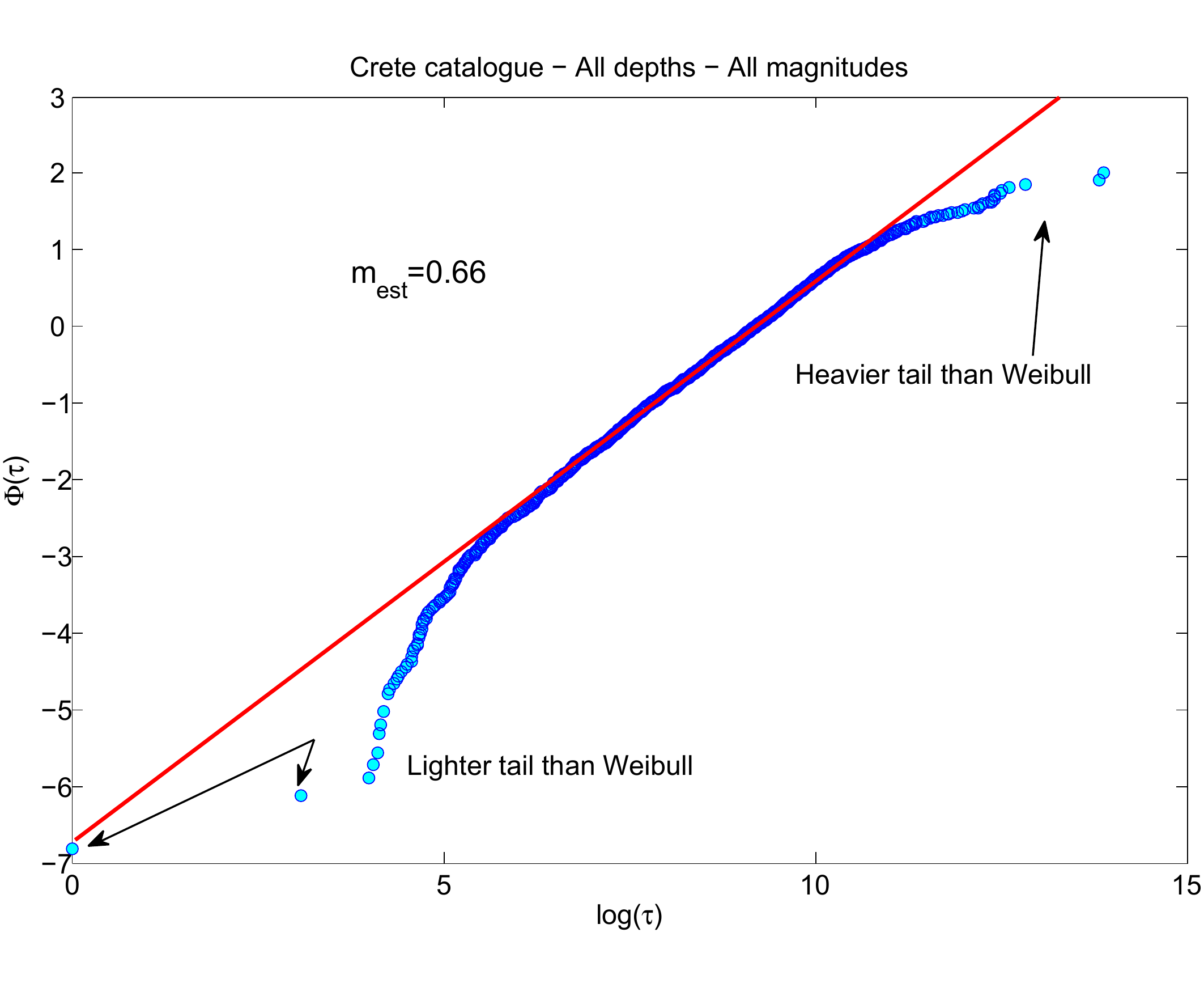}
\caption{Weibull plot of CSS interevent times.  The straight line indicates perfect Weibull dependence;
$m_{est}:=\hat{m}$ is the maximum likelihood estimate of the Weibull modulus from the data. }
\label{fig:Crete_wbl}
\end{figure}

 The Weibull plot for CSS is shown in Fig.~\ref{fig:Crete_wbl}.
The slope of the best-fit (minimum least squares) straight line is $\hat{m}=0.66$. Note that based on the analysis in~\ref{ssec:crust-strength},
since $\hat{m}<1$ we do not expect significant modification of the Weibull due to the depth dependence of $S_s$. Nevertheless, the lower tail of $\hat{\Phi}(\tau)$  is lighter than the Weibull while its upper tail is heavier. The lower tail begins roughly at $\hat{\Phi}(\tau_1) \approx -2.5$ and the upper tail near $\hat{\Phi}(\tau_2) \approx 1.4$. By inverting Eq.~\eqref{eq:Phi}, we determine that $\hat{F}_{\tau}(\tau_1) \approx 0.08$ and
$\hat{F}_{\tau}(\tau_2) \approx 0.98$.
The lower-tail behavior, which involves about $8\%$ of the data, can be attributed to unresolved events (i.e.,
with magnitudes below the magnitude of completeness).
The upper-tail behavior, which involves around $2\%$ of the data,  can be partly explained due to the same effect, since failure to observe certain seismic events leads inadvertently to  interevent times that are higher than in reality. The discrepancy could also be due to genuine departure  of the upper ITD tail from the Weibull behavior, since the CSS data set involves a system of faults with a composite crustal strength distribution, c.f. Eq.~\eqref{eq:CDF-fault-system}.

\subsection{Analysis of CSS interevent times for events above ${M}_{L,c}$}

\begin{figure*}
\centering
\subfloat[CSS interevent times]{\label{fig:Crete_wbl_Mwc}\includegraphics[width=0.47\textwidth]{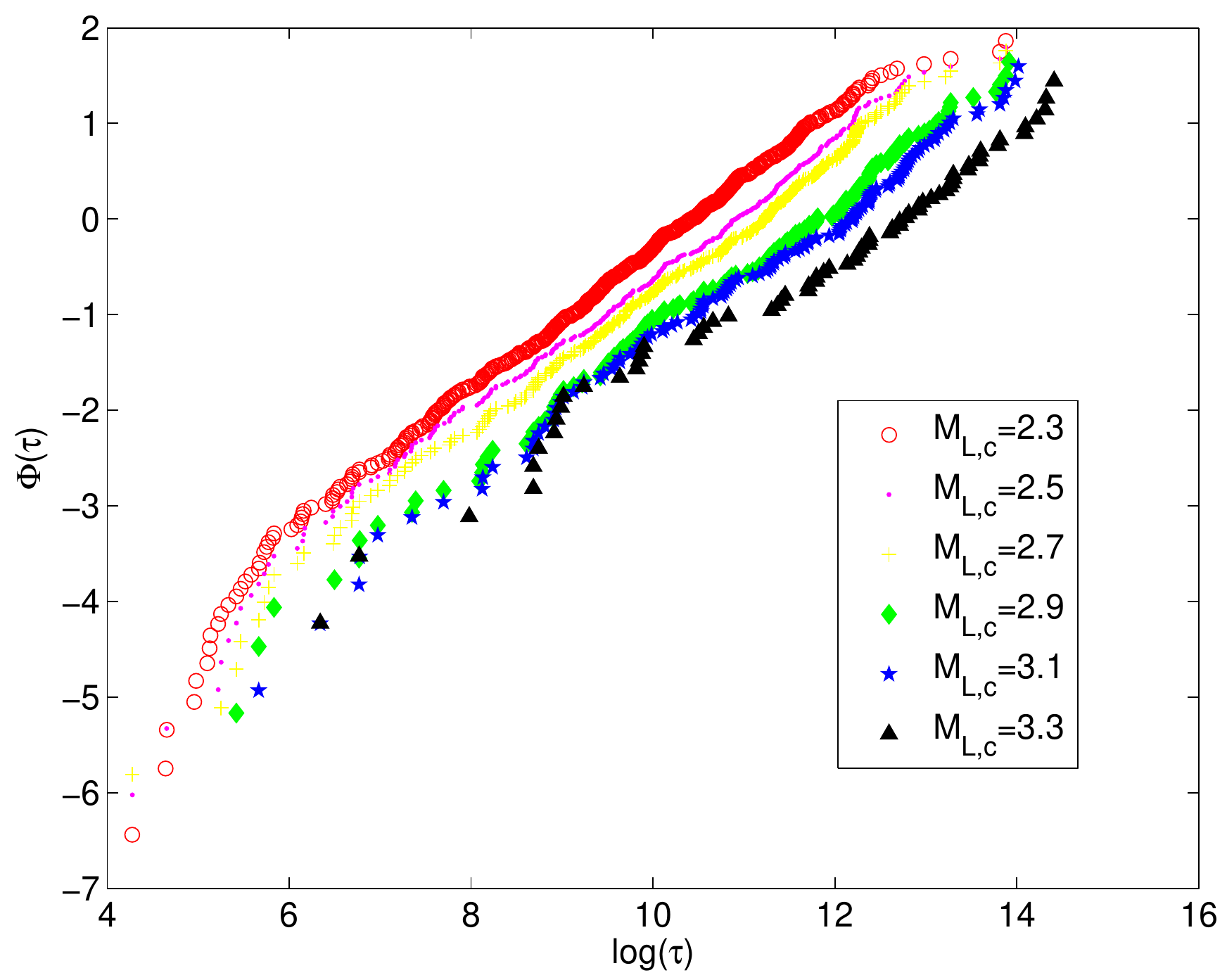}}
\subfloat[CSS time scales]{\label{fig:Crete_RIchar_Mwc}\includegraphics[width=0.47\textwidth]{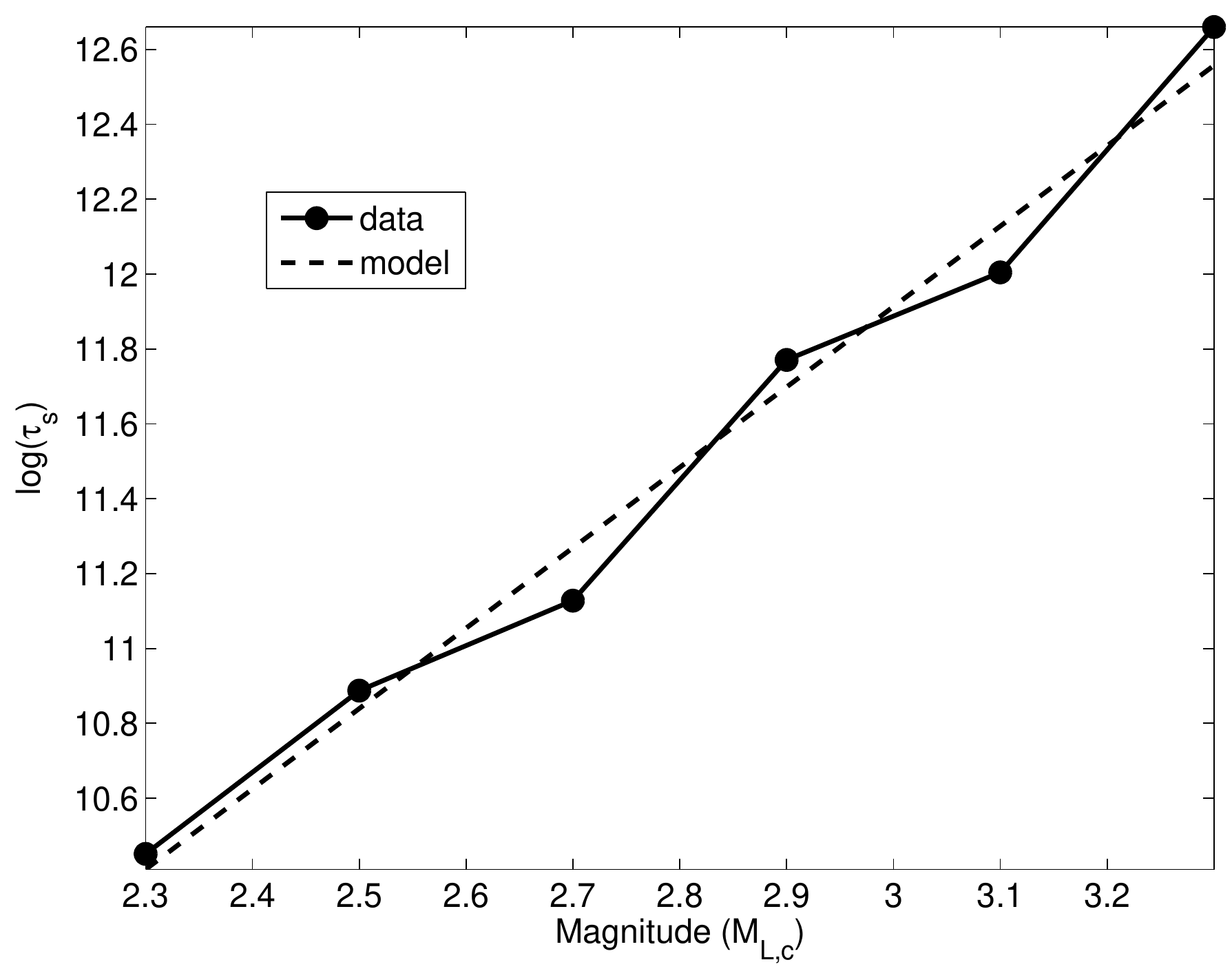}}
\caption{(a) Weibull plot of CSS interevent times for different magnitude thresholds. (b)
Semilog plot of empirical (circles) and theoretical -according to Eq.~\eqref{eq:taus-scaling}- (broken line) Weibull time scale versus ${M}_{L,c}$. }
\label{fig:CSS_with_cutoff}
\end{figure*}

Below, we focus on interevent times between events above  thresholds that exceed the magnitude of completeness.
The resulting sequences are considered to be complete since they do not suffer from resolution issues.
We show the Weibull plots for different magnitude thresholds in Fig.~\ref{fig:Crete_wbl_Mwc}.  The $\hat{\Phi}(\tau;M>{M}_{L,c})$  curves show deviations from the Weibull in both the lower and upper tails. The curvature of the upper tail appears to reverse its sign as ${M}_{L,c}$ changes from 2.7 to 2.9.  Nevertheless, the Weibull dependence remains a useful first approximation. In particular, for all the thresholds considered,                                                                                                                                                                                                                                                                                                                                                                                                                                                                                                                                                                                                                                                                                                                                                                                                                                                                                                                                                                                                                                                                                                                              the Kolmogorov - Smirnov test does not reject the null hypothesis that the CDF is the Weibull with the respective best-fit (maximum likelihood) parameters at significance level $5\%$.

In Table~\ref{tab:Weibull-par} we list the Weibull ITD statistics using different magnitude thresholds.
Based on the confidence intervals for $\hat{m}$ in Table~\ref{tab:Weibull-par}, there is no significant variation of $\hat{m}$ with ${M}_{L,c}$. In contrast, $\hat{{\rm\tau}}_s$ increases with ${M}_{L,c}$ as intuitively expected.
In Fig.~\ref{fig:Crete_RIchar_Mwc} we plot $\log(\hat{\tau}_s)$ versus ${M}_{L,c} \in [2.3, 3.3]$.
The best-fit (minimum least squares) line passing through the data is   $\log(\hat{\tau}_s) = A - \rho_{M}\,{M}_{L,c} $ with
$A=5.47, \rho_{M}=2.15$,  while the
$95\%$ confidence interval is $[1.76, 2.54].$  The value of $\rho_{M}$ predicted by the analysis in~\ref{ssec:mag-dep}, is $\rho_{M}^{\ast} = b\, \log(10) \approx 1.96$ in good agreement  with the experimental $\rho_{M}$.

\begin{table}[htb]
\caption{Number of events $(N_{>})$ with magnitude above  ${M}_{L,c}$  (2.3 - 3.3 ${M}_{L}$),
 ITD Weibull parameters $\hat{m}$ and $\hat{\tau}_s$ and their $95\%$ confidence intervals,
 $[m_1,m_2]$ and $[\tau_{s,1}, \tau_{s,2}]$ respectively. Weibull modulus is rounded to the second decimal place, while $\hat{\tau}_s$ is rounded to the third
 most important digit.}
\begin{center}
\begin{ruledtabular}
\begin{tabular}{c c c c c c}
 ${M}_{L,c} $ &  $N_{>}$  & $\hat{m}$ & $\hat{\tau}_s$ (sec) & $[m_1,m_2]$ & $[\tau_{s,1}, \tau_{s,2}]$ \\
  \hline
 2.3 & 629 & 0.67 & $3.46\, 10^4$ & [0.66, 0.74] & $[3.07 \, 10^4,  3.89\, 10^4]$\\
 2.5 & 415 & 0.70 & $5.35\, 10^4$ & [0.65, 0.75] & $[4.63 \, 10^4,  6.18\, 10^4]$ \\
 2.7 & 335 & 0.72 & $6.80\, 10^4$ & [0.66, 0.78] & $[5.81 \, 10^4,  7.97\, 10^4]$ \\
 2.9 & 177 & 0.71 &$ 1.25 \,10^5$ & [0.63, 0.80] & $[1.04 \, 10^5,  1.62\, 10^5]$ \\
 3.1 & 140 & 0.70 & $1.63 \, 10^5$ & [0.62, 0.80] & $[1.27 \, 10^5,  2.10\, 10^5]$ \\
 3.3 & 70  & 0.67 & $3.14 \, 10^5$ & [0.55, 0.81] & $[2.17 \, 10^5,  4.56\, 10^5]$ \\
\end{tabular}
\end{ruledtabular}
\end{center}
\label{tab:Weibull-par}
\end{table}

\section{Conclusions}
\label{sec:conclusions}
We propose a stochastic stick - slip  model for the earthquake interevent times distribution that is based on the evolution of shear stress in the Earth's crust
and the assumption that the crust strength
 follows the Weibull distribution, as  is common for brittle materials.
The current model differs from statistical approaches that test various empirical distributions,
from universal scaling laws based on the concept of criticality~\cite{Bak02,Corral03,Corral04,Corral06a,Corral06b}, from approaches rooted in probability theory~\cite{Santhanam08}, and from stochastic branching models~\cite{Saichev07} that incorporate by construction the Gutenberg - Richter and Omori laws. Our model is applicable to tectonic earthquakes that are responsible for the majority of the global earthquake activity. The interevent times distribution depends on both the crustal strength  and the stress accumulation models. We show that the commonly used Weibull (including the exponential) and log-Weibull interevent times distributions  are derived  from the stochastic stick - slip model  using, respectively, power-law (including linear) and logarithmic dependence of the stress accumulation function on time. We focus on power-law stress accumulation, which leads from Weibull strength statistics to Weibull interevent times. We also demonstrate that deviations from the Weibull dependence arise due to limited resolution (i.e., sampling of events exceeding a finite magnitude threshold), and random fluctuations of the accumulation rate.

We also derive the scaling relation~\eqref{eq:taus-scaling} that links the magnitude threshold
 with the scale of the respective Weibull interevent times distribution. This relation is based on the assumption that the Weibull model is an acceptable approximation of the ITD at finite magnitude thresholds (i.e., neglecting the lower-tail deviations) and on the Gutenberg - Richter law.  The utility of the scaling relation is greater if the Weibull modulus can be assumed to remain relatively constant as the magnitude threshold changes. Then, this relation can be used to infer the Weibull time scales for larger-magnitude  events based on the time scales of smaller-magnitude, more frequent, events.

In future work we will investigate dynamic stick - slip models with controlled stress accumulation rates
that will allow simulating seismic events. We will also focus on the deviations from the Weibull distribution in the upper tail. Another direction that will be pursued is the interplay
 between empirical scaling laws of fault and earthquake parameters
(e.g., fault size distribution, seismic stress drop versus fault size and earthquake magnitude) and the ITD of a fault system. In particular, a model for the behavior of a fault system should involve an average over the sizes and characteristic scales of the faults it comprises.  Such an average should account for the empirical statistical facts that govern the fault population.
While we believe that the interplay between the fault strength and the stress accumulation function is a
universal factor controlling the ITD, the specific stress accumulation scenarios proposed and  investigated herein  do not  exhaust the possible functional forms of stress accumulation. In particular, seismic events driven by  invasion of fluids in a fault system may be generated by more local and erratic stress accumulation patterns.

\begin{acknowledgments}
This research was partly supported by a Marie Curie International
Incoming Fellowship within the 7th European Community Framework
Programme under contract no. PIIF-GA-2009-235931. Seismic data for
Crete were kindly provided by D. Becker, Institute of Geophysics, Hamburg
University, Germany. We also thank Dr. R. Robinson (GNS Science) for helpful discussion on
the reference~\cite{Robinson09}.
\end{acknowledgments}

\appendix
\section{Effective strength distribution for linear depth dependence of Weibull scale parameter }
\label{App:renorm-weibull}

We assume that the strength scale depends on the depth as $S_{s}(h)= S_{0} + q\,h.$
Then, using the
variable transformation $z=\left( {S_{0} + q\,h} \right)^{-m_{s}}$, and
$ dz = - m_{s}\, q\, (S_{0} + q\,h)^{-(m_{s}+1)} \, dh= - {m_{s}\, q} \, z^{1 +1/m_{s}} \, dh,$ i.e.,
$ dh = - \frac{dz}{m_{s} \, q } \, z^{-(1 +1/m_{s})} $,
the integral representing the survival function in Eq.~\eqref{eq:CDF-fault-eff-strength-int} is expressed as follows
 \begin{equation*}
R_{\rm S}^{\ast}(s)=  \frac{1}{m_{s} \, q\, (h_2 - h_1) }\int_{z_{2}}^{z_{1}} dz \, z^{-(1 +1/m_{s})} \, e^{- z S^{m_s}},
\end{equation*}
where $z_{i}= (S_{0} + q\,h_{i})^{-m_{s}}, i=1,2.$ We now introduce the following definitions:
$2\delta h = h_2 - h_1$, $\bar{S} = \frac{S_{1} +S_{2}}{2},$ and the dimensionless variables:
$ s = S/\bar{S}$,  $\lambda_s= q \, \delta h/\bar{S},$ $ u =z \bar{S}^{m_{s}},$  and $u_{1,2} = \left( \frac{1}{1 \mp \lambda_s } \right)^{m_{s}}$, in terms of which the integral takes the form of Eq.~\eqref{eq:CDF-fault-dav-strength}.

  \subsection{Evaluation of $R_{\rm S}^{\ast}(s)$ for $m_s \ge 1$}
  We evaluate the integral in Eq.~\eqref{eq:CDF-fault-dav-strength} using integration by parts,
  using the survival function $ R_{\rm S}^{\ast}(s)= 1 -  F_{\rm S}^{\ast}(s)$. This leads to
     \begin{align*}
  \label{eq:survival-mgt1}
 R_{\rm S}^{\ast}(s) & = \frac{-1}{2  \, \lambda_s }
 \int_{u_{2}}^{u_{1}} du \, (u^{-1/m_{s}})' \, e^{- b\,u} \\
 & = \frac{-1}{2  \, \lambda_s }  \left[ \left. (u^{-1/m_{s}}\,  e^{- b\,u})\right|_{u_2}^{u_1}
 + b\, \int_{u_{2}}^{u_{1}} du \, u^{-1/m_{s}} \, e^{- b\,u} \right].
\end{align*}
The term that originates from the boundary points (recalling that $b=s^{m_s}$) is equal to
\[
 \left[ \frac{(1+\lambda_s)}{2  \, \lambda_s}\, e^{-\left( \frac{s}{1+\lambda_s} \right)^{m_s}} -
\frac{(1 - \lambda_s)}{2  \, \lambda_s}\, e^{-\left( \frac{s}{1 - \lambda_s} \right)^{m_s}} \right].
\]
For the remaining integral, we use the variable transformation $z=bu$, in terms of which we obtain
\begin{align*}
& b\, \int_{u_{2}}^{u_{1}} du \, u^{-1/m_{s}} \, e^{- b\,u} = b^{1/m_s} \, \int_{z_{2}}^{z_{1}} dz \, z^{-1/m_{s}} \, e^{- z}\\
& = b^{1/m_s} \, \left[ \gamma(1- 1/m_{s}, z_{1}) -   \gamma(1- 1/m_{s}, z_{2})    \right],
\end{align*}
 where $z_{1,2} = \left( \frac{s}{1  \mp \lambda_s } \right)^{m_s}$ and
 $\gamma(\alpha,x)$ is the \emph{lower incomplete gamma function} defined by
 $\gamma(\alpha,x) = \int_{0}^{x} du \, u^{\alpha-1} \, e^{-u}.$
 Hence, collecting the relevant terms we obtain
 \begin{align}
 \label{eq:survival_mgt1}
 R_{\rm S}^{\ast}(s) & = \left[ \frac{(1+\lambda_s)}{2  \, \lambda_s}\, e^{-\left( \frac{s}{1+\lambda_s} \right)^{m_s}} -
\frac{(1 - \lambda_s)}{2  \, \lambda_s}\, e^{-\left( \frac{s}{1 - \lambda_s} \right)^{m_s}} \right]
\nonumber \\
& - \frac{s}{2\lambda_s}\,  \left[ \gamma\left(1- \frac{1}{m_{s}},  \frac{s^{m_s}}{( 1  - \lambda_s)^{m_s}}\right) \right.  \nonumber \\
 & \left.  \quad  \quad \quad  -\gamma\left(1- \frac{1}{m_{s}},  \frac{s^{m_s}}{(1 + \lambda_s )^{m_s}} \right)     \right].
 \end{align}
The absolute difference between the CDF that is based on the numerical integration of Eq.~\eqref{eq:CDF-fault-dav-strength} and that obtained from the explicit solution~\eqref{eq:survival_mgt1} is less than $1.4 \times 10^{-8}$ over the parameter space of
 Fig.~\ref{fig:Weibull_correction}.

 \subsection{Evaluation of $R_{\rm S}^{\ast}(s)$ for $m_s <1$}

For $m_s <1$ partial integration is not as efficient as for $m_s > 1$,  because the incomplete gamma function is not defined for $1- \frac{1}{m_{s}} <0$; an extension of the gamma function is possible by means of partial integration, which reduces the incomplete gamma function of a negative argument to  one with a positive argument. Nevertheless, for $m_s <<1$ it follows that $1- \frac{1}{m_{s}} <<-1$, and repeated
partial integrations are required.   Alternatively, a series expansion of the integral in Eq.~\eqref{eq:CDF-fault-dav-strength} is obtained by means of the Taylor expansion of the exponential function $e^{-b\,u}= \sum_{n=0}^{\infty} (-1)^{n}\frac{(b\, u)^n}{n!}$ (where
$b = s^{m_s}$). The
survival function is then given by
 \begin{align}
  \label{eq:R_mlt1}
 & R_{\rm S}^{\ast}(s)  =  \frac{1}{2 m_{s} \, \lambda_s } \sum_{n=0}^{\infty}
 \frac{(- 1)^{n} \, s^{n \, m_{s}}}{n!} \int_{u_{2}}^{u_{1}} du \, u^{n-(1 +1/m_{s})} \nonumber \\
 & = \frac{1}{2  \, \lambda_s } \sum_{n=0}^{\infty}
 \frac{(- 1)^{n} \, s^{n\, m_{s}}}{n!} \frac{ \left( 1 - \lambda_s  \right)^{1 -n\, m_{s} } - \left( 1 + \lambda_s  \right)^{1 -n\, m_{s} } }{n\, m_{s}-1}.
\end{align}
The above alternating series can be expressed as $ R_{\rm S}^{\ast}(s) = \sum_{n=0}^{\infty}
 (- 1)^{n} \, a_{n},$ where
 \begin{equation}
 \label{eq:an}
 a_{n} = \frac{s^{n\, m_{s}}}{ n!} \frac{ \left( 1 - \lambda_s  \right)^{1 -n\, m_{s} } - \left( 1 + \lambda_s  \right)^{1 -n\, m_{s} } }{2  \, \lambda_s \, ( n\, m_{s}-1) }.
 \end{equation}

\begin{figure}
\centering
\includegraphics[width=0.47\textwidth]{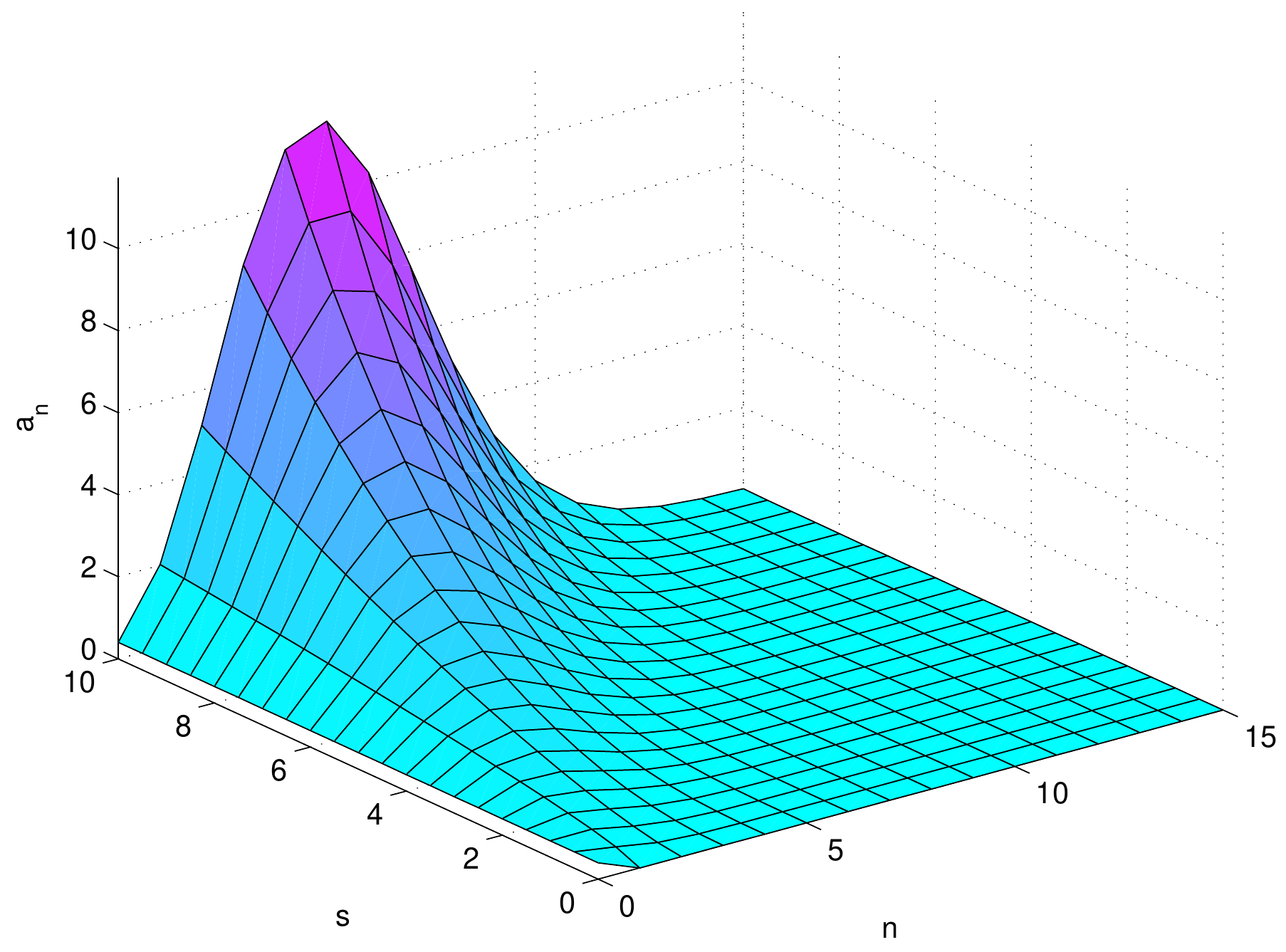}
\caption{Dependence of the terms $a_{n}$ in Eq.~\eqref{eq:an}, on $n=0,1,\ldots 15$ and  $s=0, 0.5,\ldots 10$, for parameter values $\lambda_s=0.2$, and $m_{s}=0.7$.}
\label{fig:Series_terms}
\end{figure}

 The sequence $a_{n}$ has a maximum for an integer  $n^{\ast}$ which depends on $s,  \lambda_s, m_s$, as shown
 graphically in Fig.~\ref{fig:Series_terms}.
 Fixing these three parameters, we can write $ R_{\rm S}^{\ast}(s)  = R_{\rm S}^{(f)}(s) + R_{\rm S}^{(in)}(s)$
  where $R_{\rm S}^{(f)}(s)$ is the  finite series
 $R_{\rm S}^{(f)}(s) = \sum_{n=0}^{n^{\ast}}(- 1)^{n} \, a_{n}$ and
 $R_{\rm S}^{(in)}(s) =  (-1)^{n^{\ast}} \sum_{n=1}^{\infty}(- 1)^{n} \, {a'}_{n} ,$
 where ${a'}_{n} = a_{n^{\ast}+n} $.
 For $n > 1$ it holds that ${a'}_{n} \le {a'}_{n+1}.$
 Based on the alternating series test, $R_{\rm S}^{(in)}(s)$ converges. In addition, if
  $R_{\rm S}^{(in)}(s)$ is truncated after $M$ terms, the absolute value of the remainder is less
  than $a_{M+1}$.
  The absolute difference between the CDF that is based on the numerical integration of Eq.~\eqref{eq:CDF-fault-dav-strength} and that obtained from the series~\eqref{eq:R_mlt1} truncated
  at $n=100$ is less than $2.2 \times 10^{-9}$ over the parameter space of
 Fig.~\ref{fig:Weibull_correction}.

\newpage

\bibliography{DTH_VM_ITD_2011}
\newpage


\end{document}